\documentclass[a4paper,11pt]{article}
\usepackage{jheppub} % for details on the use of the package, please see the JINST-author-manual
\usepackage[labelfont=bf]{caption}
\usepackage{lineno}
\usepackage{slashed}
\usepackage{booktabs}
\usepackage{arydshln}
\usepackage{amssymb}
\usepackage{amsthm}
\usepackage{cancel}
\usepackage{slashed}
\usepackage{bm}
\usepackage{dsfont}
\usepackage{amsmath}
\usepackage{multicol}
\usepackage{multirow}
\usepackage{xcolor}
\usepackage{color}
\definecolor{myblueold}{RGB}{65,105,225}
\definecolor{myblue}{RGB}{0, 71, 171}
\definecolor{mahogany}{RGB}{156, 0, 0}
\definecolor{mygray}{RGB}{145,145,145}
\definecolor{mygreen}{RGB}{34,139,34}
\definecolor{mybrown}{RGB}{110, 38, 14}
\usepackage{subcaption}

\usepackage{hyperref}

\makeatletter
\def\hlinewd#1{%
	\noalign{\ifnum0=`}\fi\hrule \@height #1 %
	\futurelet\reserved@a\@xhline}
\makeatother

\newcounter{RSQ}

\newcounter{DFQ}

\newcommand{\df}[1]{\textcolor{myblueold}{ #1}}

\allowdisplaybreaks

\def\fn{\footnote}
\def\be{\begin{equation}}
\def\ee{\end{equation}}
\def\ba{\begin{alignedat}}
\def\ea{\end{alignedat}}
\def\bea{\begin{eqnarray}}
\def\eea{\end{eqnarray}}
\newcommand{\bs}{\begin{subequations}}
\newcommand{\es}{\end{subequations}}
\usepackage{sectsty}
\sectionfont{\Large}
\subsectionfont{\large}
\subsubsectionfont{\normalsize}

\renewcommand{\eqref}[1]{(\ref{#1})}

\newcommand{\ali}[1]{\begin{align}#1\end{align}}

\def\II{\textcolor{mahogany}{\rm(II)}}
\def\III{\textcolor{mahogany}{\rm(III)}}
\def\IV{\textcolor{mahogany}{\rm(IV)}}
\def\V{\textcolor{mahogany}{\rm(V)}}

\def\zz{\textcolor{mygray}{(0)}}
\def\zzzz{\textcolor{mygray}{(00)}}

\newcommand{\ar}[1]{\textcolor{mybrown}{(#1)}}

\def\h{\textcolor{myblue}{(h)}}
\def\sh{\textcolor{myblue}{(sh)}}
\def\hc{\textcolor{myblue}{(hc)}}
\def\p{\textcolor{myblue}{(p)}}
\def\s{\textcolor{myblue}{(s)}}
\def\c{\textcolor{myblue}{(c)}}
\def\sc{\textcolor{myblue}{(sc)}}

\def\hnvar{\textcolor{myblue}{hn}}
\def\hvar{\textcolor{myblue}{h}}
\def\shvar{\textcolor{myblue}{sh}}
\def\svar{\textcolor{myblue}{s}}

\def\scvar{\textcolor{myblue}{sc}}

\title{\boldmath EFT approach to the endpoint of muon decay-in-orbit}

\author[a]{Duarte Fontes,}
\author[b]{Robert Szafron}
\affiliation[a]{Institute for Theoretical Physics,
Karlsruhe Institute of Technology,
76128 Karlsruhe, Germany}
\affiliation[b]{Brookhaven National Laboratory, Upton, NY, U.S.A.}

\date{} % This removes the date

\emailAdd{duarte.fontes@kit.edu}
\emailAdd{rszafron@bnl.gov}

\abstract{
As upcoming experiments aim to probe muon conversion with unprecedented precision, equally precise theoretical predictions are crucial to maximize discovery potential. This applies not only to the new physics signal, muon-electron conversion, but also to its only irreducible background, muon decay-in-orbit (DIO) near the endpoint. Accurate computation of higher-order corrections in bound states is a long-standing challenge due to the difficulty of systematically organizing contributions. In previous work, we developed an Effective Field Theory framework to address this issue and applied it to muon conversion. Here, we extend this approach to the DIO endpoint, a more complex problem due to the presence of a neutrino-antineutrino pair in the final state. We present the most precise prediction to date of the background spectrum relevant for future muon conversion searches, achieving next-to-leading logarithmic prime accuracy for QED corrections.
}

\makeatletter
\gdef\@fpheader{}
\makeatother
\begin{document}

\begin{flushright}
KA-TP-9-2025\\
P3H-25-042
\end{flushright}
\vspace*{1cm}

\maketitle
\flushbottom

\section{Introduction}
\label{sec:intro}

Muon conversion, the process in which a muon converts to an electron in the vicinity of a nucleus, remains to be observed. It provides one of the best limits on charged lepton flavor violation (CLFV), expressed as $R_{\mu e} < 7 \times 10^{-13}$ \cite{SINDRUMII:2006dvw}. Since CLFV is in practice absent in the Standard Model (SM) due to the smallness of neutrino masses, a detection of muon conversion would provide an indisputable sign of physics beyond the SM (BSM).

Such detection might happen in the near future, as upcoming CLFV experiments are expected to achieve a four-order-of-magnitude reduction in the bound on 
$R_{\mu e}$ through improved sensitivity \cite{Mu2e:2014fns, Diociaiuti:2024stz,COMET:2018auw, Fujii:2023vgo,Mu2e-II:2022blh}. This remarkable experimental improvement requires precise theoretical predictions for both the signal and its Standard Model (SM) backgrounds. It turns out that there is only one non-reducible background: muon decay-in-orbit (DIO), consisting of a muon decay in the vicinity of a nucleus. More specifically, it is the region of the muon DIO spectrum near the electron-energy endpoint (for short, endpoint DIO, or eDIO)
%--- where the electron energy is similar to the muon mass --- 
that forms the dominant background for muon conversion searches.
To optimize the signal-to-background ratio and so enhance the chances of observing muon conversion, upcoming experiments require precise theoretical predictions for the shape of each rate (i.e. for the electron-energy spectra of both muon conversion and eDIO).
Since that shape is dictated by the QED corrections (i.e. higher-order effects in the fine structure constant $\alpha$), it is of the utmost importance to have a proper theoretical prediction for the latter.

However, QED corrections in bound states are notoriously challenging. They demand advanced techniques to rigorously define the relevant operators, consistently organize the various contributions, and systematically avoid double counting. In addition, they are often dominated by large logarithms arising from the hierarchy of energy scales, which spoil the convergence of the perturbative expansion. We addressed these challenges in a previous publication \cite{Fontes:2024yvw}, where we developed a framework that cleanly separates the multiple scales at play using Effective Field Theory (EFT) techniques. This approach allows different types of physics to be treated systematically, as muon conversion and eDIO simultaneously involve a heavy nucleus, a non-relativistic muon, an energetic and massive electron, and soft real radiation. More specifically, the nucleus requires Heavy Quark Effective Theory (HQET) \cite{Isgur:1989vq, Isgur:1990yhj, Neubert:1993mb, Manohar:1997qy, Manohar:2000dt},
the muon calls for both Non-Relativistic QED (NRQED) \cite{Caswell:1985ui, Kinoshita:1995mt, Paz:2015uga} and potential NRQED (pNRQED) \cite{Pineda:1997bj, Pineda:1997ie, Brambilla:1999xf, Beneke:1999qg, Beneke:1998jj}, the energetic electron entails Soft-Collinear Effective Theory (SCET) I and II \cite{Bauer:2000ew, Bauer:2000yr, Bauer:2001ct, Bauer:2001yt, Beneke:2002ph, Beneke:2002ni}, and the soft real radiation coupled to massive energetic particles requires boosted HQET (bHQET)~\cite{Fleming:2007qr, Fleming:2007xt}.

Although this EFT framework applies to both muon conversion and eDIO (since the hierarchy of the relevant physical scales is the same in both), ref.~\cite{Fontes:2024yvw} focused on the former and on $\mathcal{O}(\alpha)$ corrections to that process. In this paper, we do the same for eDIO, which is complicated by the presence of the neutrino-antineutrino pair in the final state. Several works have discussed theoretical improvements for eDIO
\cite{Shanker:1979ap,Shanker:1981mi,Shanker:1996rz,Czarnecki:2011mx,Szafron:2015kja,Szafron:2016cbv,Szafron:2017guu}. 
In particular, a great deal of recent efforts has been devoted to understanding the impact of nuclear effects on eDIO \cite{Heeck:2021adh,Kaygorodov:2025yag}. Radiative effects pose a complementary problem that has received much less attention in the literature. Notably, ref.~\cite{Szafron:2015kja} calculated $\mathcal{O}(\alpha)$ corrections for eDIO, but lacked a formalism capable of addressing the complexities associated with higher-order corrections in bound states, as discussed above. By filling in this gap, our work provides not only the most precise shape of the rate for eDIO, but also the foundations to consistently improve the rate even more. 

The paper is organized as follows. We start by discussing basic aspects of eDIO in section \ref{sec:Basics}, after which we turn to the EFT framework in section \ref{sec:EFT}. There, we build on the concepts and techniques presented in ref.~\cite{Fontes:2024yvw}, emphasizing the elements unique to eDIO.
In section \ref{sec:Factorization} we derive a leading power factorization theorem for the eDIO rate, which allows us to show the most accurate predictions for its shape in section \ref{sec:Results}. After presenting our conclusions and outlook in section \ref{sec:Conclusions}, we provide technical details in the appendix.

\section{Overview of the muon decay-in-orbit endpoint}
\label{sec:Basics}

\begin{figure}[h!]
\centering
\begin{minipage}[c]{0.35\textwidth}
  \centering
  \raisebox{10mm}{\includegraphics[width=\textwidth]{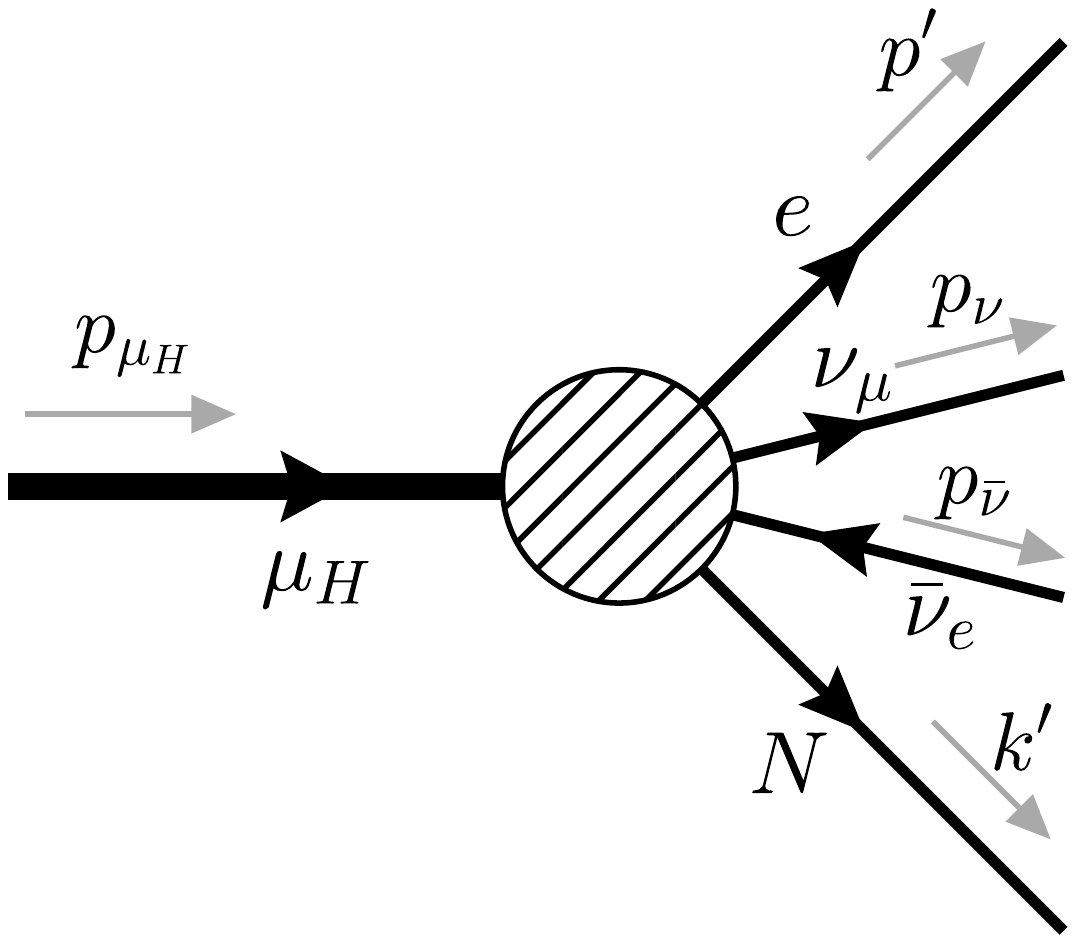}}
\end{minipage}
\hspace{8mm}
\begin{minipage}[c]{0.40\textwidth}
  \centering
  \includegraphics[width=\textwidth]{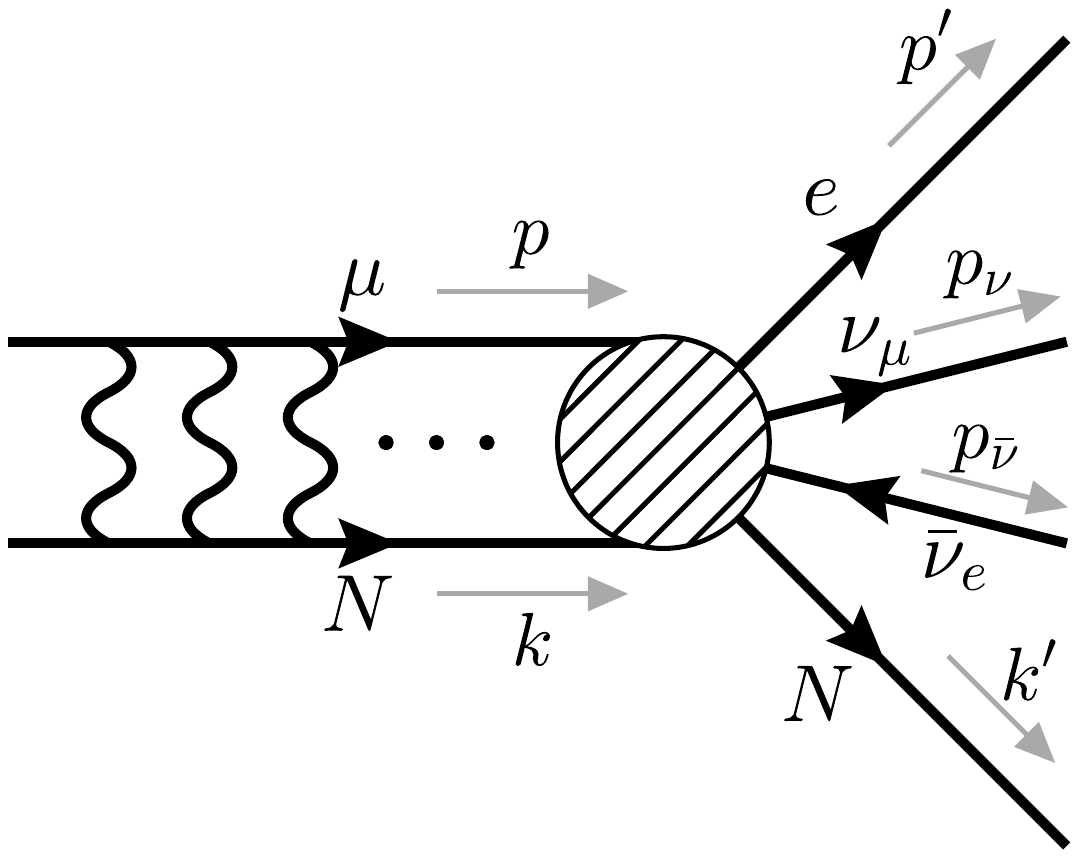}
\end{minipage}
\vspace{-35mm}
\caption{Bound muon decay. The left diagram shows the bound state $\mu_H$ as a single field, whereas the right diagram illustrates its composite structure. In both cases, the shaded region represents short-distance perturbative interactions, distinct from the long-distance Coulomb exchanges. The possible final state radiation $\mathcal{X}$ is omitted. See text for details.}
\label{fig:DIO-kinematics}
\end{figure}

Muon DIO is the decay of a muonic atom via the standard muon decay process.
%into an electron and a pair of neutrinos.
It is depicted on the left side of figure \ref{fig:DIO-kinematics}.
The muonic atom in the initial state (which we refer to as muonic hydrogen, $\mu_H$) has the total mass $M_{\mu_{H}}$, and is a bound state formed by a muon ($\mu$) with mass $m_{\mu}$ and a nucleus ($N$) with mass $M_N$ and atomic number $Z$.%
\fn{In what follows, we assume that $Z=13$ (aluminum), implying $Z \alpha \simeq 0.1$, though we keep $Z$ as a parameter. This justifies and fixes the hierarchy between the physical scales that we adopt. }
The final state includes the recoiling nucleus $N$, an energetic electron ($e$) of mass $m_e$ and energy $E_e$, a muon neutrino ($\nu_{\mu}$) and an electron antineutrino ($\bar{\nu}_e$) --- both of which we treat as massless --- and, in principle, arbitrary radiation denoted in what follows as $\mathcal{X}$ (which, near the endpoint, is necessarily soft). 

The right side of figure \ref{fig:DIO-kinematics} shows the structure of $\mu_H$. In the ground ($1s$) state, the bound muon has a binding energy $E_b$, which in the non-relativistic approximation equals $E_b = -(Z\alpha)^2 m_{\mu}/2$, and a typical velocity relative to the nucleus $v = Z\alpha$. The dots in the diagram represent an infinite number of (potential) photons exchanged between the nucleus and the muon. These exchanges form a ladder of velocity-enhanced interactions, each rung contributing a factor proportional to $Z\alpha/v$. Because they are unsuppressed, they must be summed to all orders using the Schrödinger equation.
The shaded disk represents arbitrary short-distance interactions among the various fields. Unlike the photon ladder, these interactions are perturbative, which is a consequence of considering eDIO, where $E_e \sim \mathcal{O}(m_\mu)$. In fact, the momentum transfer between the nucleus and the leptons in that case is of the order of the muon mass; such a large momentum transfer puts the intermediate states highly off-shell and thus removes the enhancement of the ladder interactions. $E_e \sim \mathcal{O}(m_\mu)$ also implies that the outgoing electron is ultra-relativistic; hence, the Coulomb interactions between the electron and the nucleus are not velocity-enhanced and so need not be resummed to all orders in $Z\alpha$.%
\fn{These effects can be easily accounted by including the electron scattering wave-function; see e.g. refs.~\cite{Berestetskii:1982qgu,Hill:2023acw,Hill:2023bfh}.}

In this paper, we are interested in the calculation of the eDIO rate in the SM, since it is this that is taken as a background in muon conversion searches.
Different approaches can be used to calculate the rate. In addition to the EFT approach --- which is systematically introduced and applied for the first time in this paper --- two other approaches have been considered in the literature. One of them, traditionally used, resorts to numerical methods both to solve the Dirac equation at leading order (LO) in $\alpha$, as well as to calculate the overlap integrals \cite{Uberall:1960zz,Haenggi:1974hp,Shanker:1981mi,Watanabe:1987su,Watanabe:1993emp,Shanker:1996rz,Czarnecki:2011mx,Heeck:2021adh}. This treatment is exact in the muon velocity $v = Z\alpha$ and makes it straightforward to include finite-nucleus-size effects (recoil corrections can only be included perturbatively \cite{Czarnecki:2011mx}). 
In principle, this approach can be extended to the Furry picture \cite{Furry:1951zz}, so as to go beyond the LO in $\alpha$. In practice, however, such computation is extremely challenging and has not yet been achieved. 

The other approach used to calculate the eDIO rate is dubbed in what follows the SC approach \cite{Szafron:2015kja}. Besides the expansion in powers of $\alpha$, it contains a double expansion. The first is around the endpoint and amounts to retaining only the leading term in $\Delta E$, defined as the small difference between $E_e$ and the maximum (i.e. endpoint) energy of the outgoing electron, $E_e^{\rm max}$. Then, the second expansion is performed, in powers of the velocity $Z\alpha$, thereby establishing a connection with traditional Feynman diagrams for scattering processes. In the SC approach, the rate can be calculated at the next-to-leading order (NLO) in $\alpha$; moreover, the calculation can be performed analytically. This was done in ref.~\cite{Szafron:2015kja} by combining the fixed order NLO result, on the one hand, with the YFS resummation \cite{Yennie:1961ad} of the soft photons, on the other.
Since it lacks systematic scale separation, the SC approach is not suitable for resummation of collinear logarithms. Furthermore, it lacks both the transparent field-theoretical definitions and the systematic improvability that characterize the EFT approach. In particular, it does not provide a clear path towards calculating higher-power corrections in $\Delta E/m_\mu$, which naturally appear in the EFT approach as power corrections. It also does not justify why YFS resummation is applicable to bound muon. 
However, it provides the foundation for the EFT treatment described in this article. 

\begin{figure}[t!]
\centering
\includegraphics[width=0.5\textwidth]{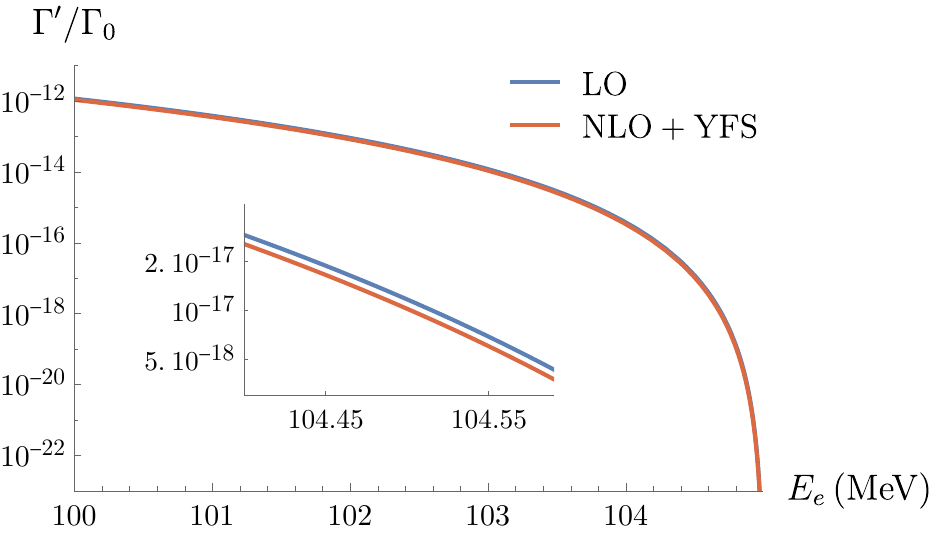}
\hspace{-4mm}
\includegraphics[width=0.5\textwidth]{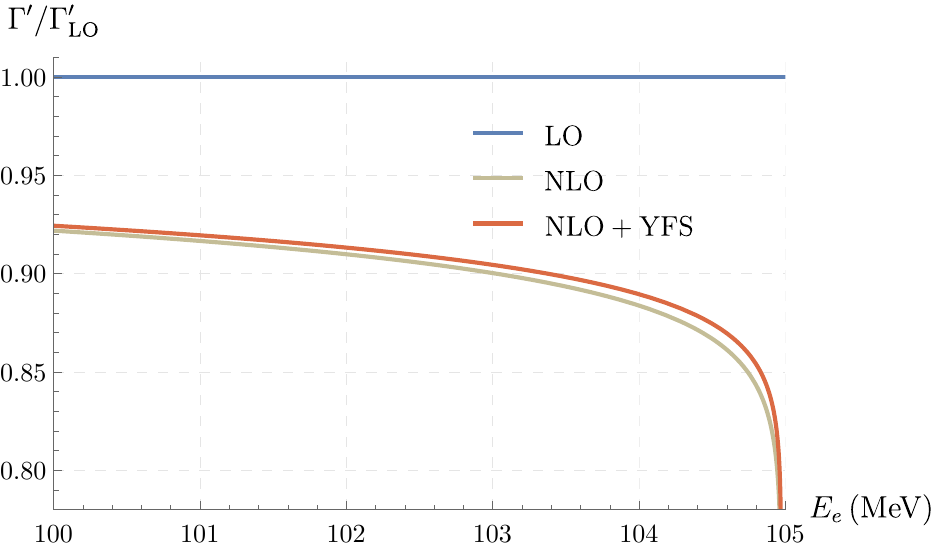}
\caption{Differential eDIO rate in the SC approach. Left panel: LO and NLO+YFS rates, normalized to $\Gamma_0$, with inset for $\Delta E \simeq m_e$. Right: LO, NLO and NLO+YFS rates normalized to LO.}
\label{fig:sc-LO-vs-NLO}
\end{figure}

Therefore, we start by reviewing the SC approach. We restrict ourselves to the results here, and leave the details to the appendix.  At LO in $\alpha$ and $Z\alpha$, we have%
\ali{
\label{eq:rate-LO}
\Gamma'_{\textrm{LO}} \equiv \dfrac{d \Gamma_{\mu_{H} \to e N \nu_{\mu} \bar{\nu}_e}^{\textrm{LO}}}{dE_e} = \dfrac{1024 \, \Gamma_0 \, Z^5 \alpha^5 \Delta E^5}{5 \pi m_{\mu}^6} + \mathcal{O}\big(\Delta E^6\big),
}
where $\Gamma_0 = G_F^2 m_{\mu}^5/\left(192 \pi^3\right)$ is the LO free muon decay rate in the limit of a massless electron.
For aluminum ($Z=13$) and $\Delta E=m_e$, $\Gamma'_{\textrm{LO}} \simeq 3.76 \times 10^{-33}$.
At NLO in $\alpha$, and still assuming aluminum, we write
\ali{
\label{eq:rate-NLO}
\frac{\Gamma'_{\textrm{NLO}}}{\Gamma'_{\textrm{LO}}}
&= 1+ \dfrac{\alpha}{\pi} \Bigg\{- \frac{26}{15} \ln \left(\frac{m_{\mu}}{m_e}\right) + \left[ 2 \ln \left(\frac{2 m_{\mu}}{m_e}\right) - 2 \right] \ln \left(\frac{\Delta E}{m_{\mu}}\right) + 6.31\Bigg\},
}
where the term 6.31 comes from the vertex corrections, vacuum polarization effects and the correction to the muon wave-function at the origin.%
\fn{In the subsequent parts of this article, we will provide the interpretation of this factor in terms of the long and short distance contributions.}
Supplementing the NLO result with YFS resummation for soft photons, we have%
\fn{Here in what follows, `LO' and `NLO' refer by default to orders in powers of $\alpha$.}
\ali{
\label{eq:rate-NLOprime}
\frac{\Gamma'_{\textrm{NLO+YFS}}}{\Gamma'_{\textrm{LO}}}
&= \left(\frac{\Delta E}{m_{\mu}}\right)^{\frac{\alpha}{\pi} \left[ 2 \ln \left(\frac{2 m_{\mu}}{m_e}\right) - 2 \right]} + \frac{\alpha}{\pi}  \bigg[6.31 - \frac{26}{15} \ln \left(\frac{m_{\mu}}{m_e}\right)\bigg].
}
We compare the differential rates at LO, NLO and NLO+YFS in figure \ref{fig:sc-LO-vs-NLO}, normalized to $\Gamma_0$ on the left panel and to $\Gamma'_{\textrm{LO}}$ on the right one. The endpoint energy is $E_e^{\rm max} := m_{\mu} - E_{\rm rec}+ E_b = 104.971$, where $E_{\rm rec}$ is the recoil energy \cite{Szafron:2016cbv}. The right panel illustrates that the NLO corrections are large (around $13\%$ in modulus for $\Delta E \simeq m_e$), a result that has its origins in the aforementioned large logarithms present in the calculation. As already suggested, this provides one of the motivations to introduce the EFT approach that follows.
%
%
%

% Eq.~(\ref{eq:rate-NLOprime}) describes large corrections (for $\Delta E=m_e$, it corresponds to a correction of $-12.4\%$ to the LO result). This is in part due to the presence of large logarithms; in fact, given $\Delta E \sim m_e \ll m_{\mu}$, all the logarithms of eq.~(\ref{eq:rate-NLO}) are large, and thus jeopardize the convergence of perturbation theory. This issue is solved by the EFT approach, to which we now turn.

Before that, we comment on our treatment of the nucleus. We follow here the approach of ref.~\cite{Fontes:2024yvw}, which we briefly summarize. We take the nucleus $N$ as a dynamical, elementary field, instead of a field composed of nucleons or quarks. This is justified by the fact that we are interested in energies at and below the muon mass scale.
In the leading approximation in the expansion around the endpoint, the nuclear charge distribution only affects the overall normalization of the eDIO spectrum; that is, it does not affect the shape of the spectrum, which is our focus here. The prescription to include finite-nucleus-size effects has been described in ref.~\cite{Szafron:2015kja}, and can be
%easily
included in the EFT framework through spatially non-local operators.

\section{EFT}
\label{sec:EFT}

As discussed in the Introduction, the EFT framework required to perform perturbative calculations in both muon conversion and eDIO was developed in ref.~\cite{Fontes:2024yvw}. The technical details are provided there and will generally not be repeated here. Instead, we summarize the essential elements, with a focus on eDIO, especially on the technical complications related to the presence of neutrinos in the final state.

\begin{figure}[t!]
%\vspace{-2mm}
\centering
\includegraphics[width=1\textwidth]{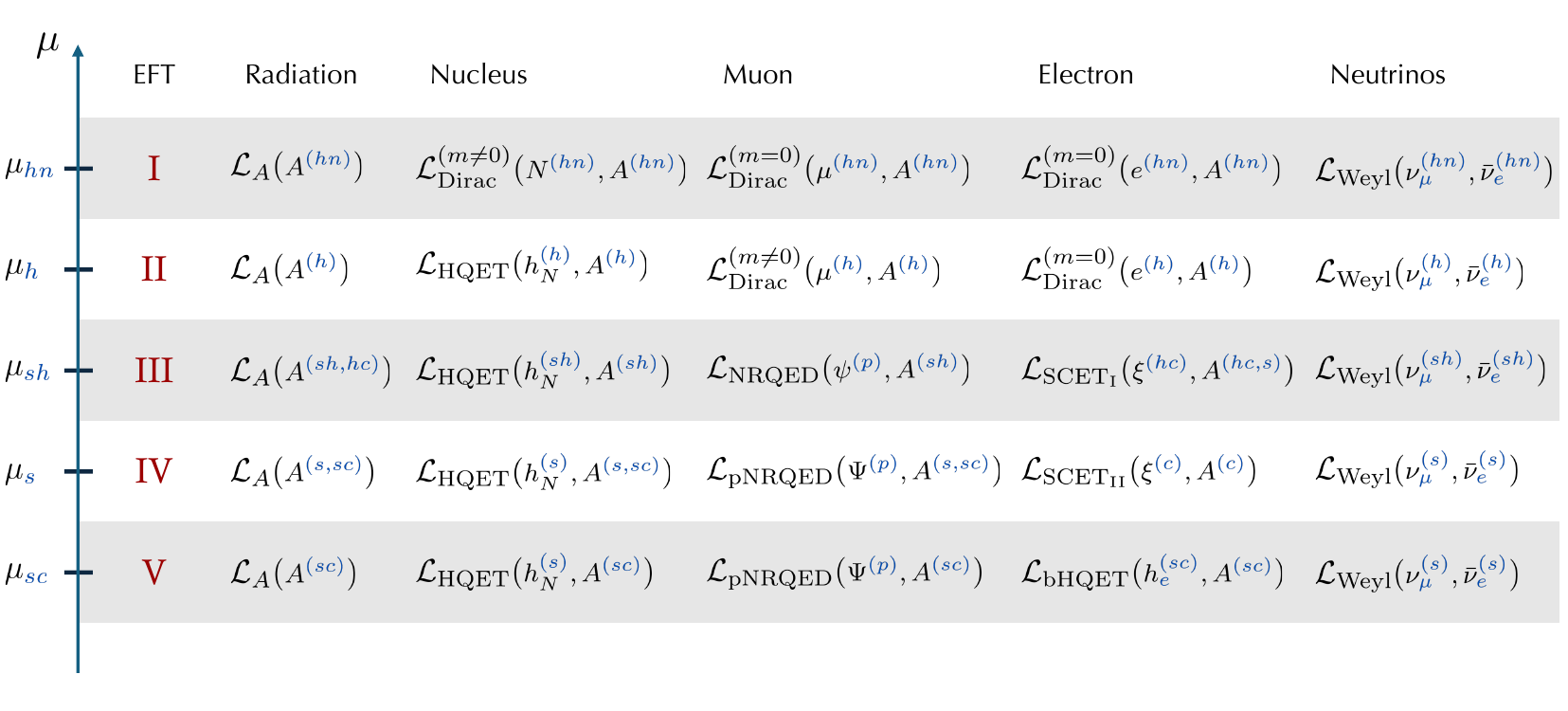}
\vspace{-10mm}
\caption{Chart of the scales relevant for muon eDIO. Each scale is associated with an EFT. See text for details.}
\label{fig:EFT-chart}
\end{figure}

The eDIO spectrum is characterized by a hierarchy of scales, 
\ali{
\label{eq:hierarchy}
M_N \gg m_\mu \sim E_e \gg Z \alpha m_{\mu} \gg (Z \alpha)^2 m_{\mu} \sim m_e \sim \Delta E, 
}
which determines the EFT framework.
%needed to obtain a systematic power expansion of the observables. 
%
That is, the framework is built to clearly separate these different scales, and allow a systematic, field-theoretical computation of the physical observables. 
It corresponds to a double expansion, the recoil and the power expansion, respectively organized by powers of the parameters
\ali{
\label{eq:scaling-def}
	\lambda_R \equiv \dfrac{m_{\mu}}{M_N},
	\qquad
	\lambda \equiv Z \alpha \sim \sqrt{\dfrac{m_e}{m_{\mu}}}.
}
In this work, we focus on leading recoil and leading power (LP) contributions. Nonetheless, the power-suppressed terms, which correspond to $Z\alpha$ and $\Delta E/m_\mu$ corrections, can be computed using the usual EFT methods.   
Given eq.~(\ref{eq:hierarchy}), the EFT framework comprises five different physical scales,
\ali{
\hspace{25mm} \textrm{hard-nuclear scale:}& \hspace{-2mm} &\mu_{\hnvar} &\sim M_N, \nonumber \\[1mm]
\hspace{25mm} \textrm{hard scale:}& \hspace{-2mm} &\mu_{\hvar} &\sim m_{\mu} \simeq E_e  = \mathcal{O}(\lambda_R \, M_N), \nonumber \\[1mm]
\hspace{25mm} \textrm{semi-hard scale:}& \hspace{-2mm} &\mu_{\shvar} &\sim Z \alpha m_{\mu} = \mathcal{O}(\lambda \, \mu_{\hvar}), \nonumber \\[1mm]
\hspace{25mm} \textrm{soft scale:}& \hspace{-2mm} &\mu_{\svar} &\sim (Z \alpha)^2 m_{\mu} \simeq m_e \simeq \Delta E = \mathcal{O}(\lambda^2 \, \mu_{\hvar}), \nonumber \\[-0.2mm]
\hspace{25mm} \textrm{soft-collinear scale:}& \hspace{-2mm} &\mu_{\scvar} &\sim m_e \frac{\Delta E}{m_{\mu}} = \mathcal{O}(\lambda^4 \, \mu_{\hvar})\nonumber,
}
and uses a different EFT at each of the five scales. Figure \ref{fig:EFT-chart} illustrates this structure, showing the Lagrangian governing the relevant dynamical fields in each EFT. This chart differs from that of ref.~\cite{Fontes:2024yvw} only by the inclusion of neutrinos. These carry negligible energy, are electrically neutral and assumed to be massless; thus, it is sufficient to describe them using the Weyl Lagrangian at all scales. 

In what follows, we subsequently describe the relevant elements of each EFT. We start with EFT II, since EFT I is not relevant for a perturbative description of eDIO \cite{Fontes:2024yvw}, and is only introduced to bridge our framework with the treatment of nuclear effects.%
\fn{We keep the EFT numbering convention of ref.~\cite{Fontes:2024yvw}.}
For each of the following EFTs, we discuss the relevant operators as well as their matching and running.

As a final note, it is worth mentioning a duality in muon DIO, related to bound \textit{vs.} unbound muons --- the simple analog of the well-known quark-hadron duality in QCD. The goal of the EFT framework that we are about to describe is the derivation of a factorization theorem. This theorem concerns the rate for eDIO, which involves a \textit{bound} state. 
However, certain objects in the theorem are defined at high scales (for example, the hard function), where the QED effects are perturbative. As such, those objects are computed with the usual Feynman rules corresponding to \textit{free} asymptotic states. This reveals the scattering process underlying the bound-state calculation, and reflects the duality between the bound muon decay and coherent scattering of muons on the nucleon.

\subsection{EFT II: \texorpdfstring{$\mu \sim \mu_{\hvar}$}{mu ≃ mus}}
EFT II is defined at the hard scale, taken to be of the order of the muon mass. The modes with higher virtuality, such as hard-nuclear modes (of the order of $M_N$), are integrated out. It follows that, at leading recoil, the nucleus is taken as infinitely heavy and is described by a static HQET field. The recoil corrections can be systematically computed using power-suppressed HQET interactions. For leptons, EFT II can be seen as the theory of weak interactions below the electroweak scale. The weak EFT Lagrangian is obtained after integrating out the scales above the muon mass. Accordingly, the muon decay is described by the renowned Fermi 4-fermion interaction,
\ali{
\mathcal{O}_{1}^{\II} &\equiv \bar{e}^{\h} \gamma_{\rho} P_L \mu^{\h} \, \bar{\nu}_{\mu}^{\h} \gamma^{\rho} P_L \nu_{e}^{\h},
}
with $P_L$ being the left-chirality projection operator. For brevity, whenever all the fields are evaluated at the same space-time point, we omit the position arguments of fields.
The Fermi Lagrangian reads 
\begin{align}
    \mathcal{L}_{\rm Fermi} = \frac{4G_F}{\sqrt{2}} \mathcal{O}_{1}^{\II} + \textrm{h.c.},
\end{align}
where $G_F$ is the matching coefficient commonly known as the Fermi constant. Since this constant does not run (i.e. it is scale independent), we take all parameters of the weak Lagrangian as defined at the hard scale.

\begin{figure}[h!]
\vspace{-5mm}
\centering
\includegraphics[width=0.73\textwidth]{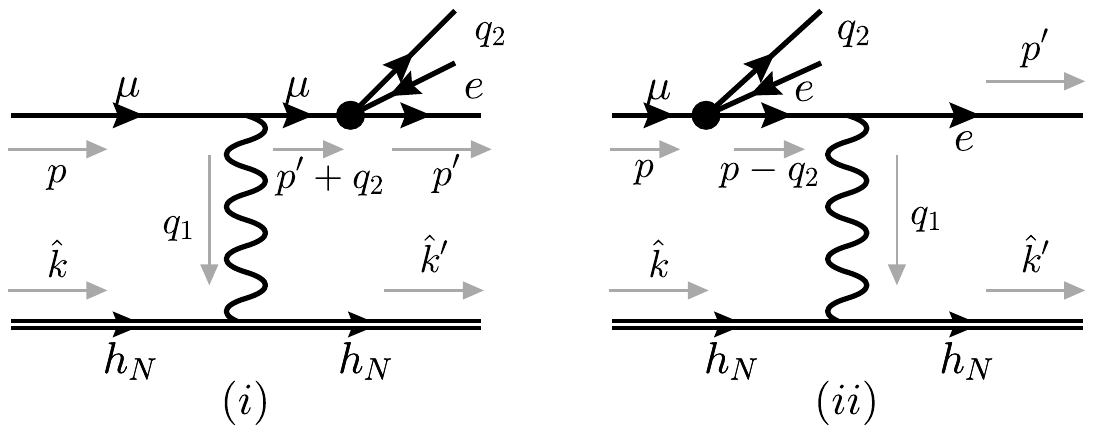}
\vspace{-100mm}
\caption{LO scattering of a muon into a nucleus with the subsequent muon decay. The black circles represent the 4-Fermi interaction. The double line represents an HQET field.}
\label{fig:DIO-LO}
\end{figure}

Anticipating the matching between EFTs II and III, we start by considering the amplitude $\mathcal{A}$ for the scattering process $\mu N \to e N \nu_{\mu} \bar{\nu}_e$.%
\fn{We recall the discussion on duality. We also note that the matching coefficients do not depend on the external states; hence, we choose the simplest states that give non-zero overlap with our operators.}
As was shown in ref.~\cite{Szafron:2015kja}, the LO contribution is associated with a single-photon exchanged between the nucleus and the charged leptons.% 
\fn{This contribution is similar to that in the spectator scattering in exclusive (semi-)hadronic decays of $B$ mesons \cite{Beneke:1999br,Beneke:2000ry,Beneke:2003pa,Neubert:2004dr,Beneke:2005gs,Beneke:2015wfa}.}
The Feynman diagrams for this contribution are depicted in fig. \ref{fig:DIO-LO}, with $h_N$ being the nucleus field.
The muon is characterized by a four-momentum $p$, 
the pair of neutrino-antineutrino by their total four-momentum $q_2$, 
and the virtual photon exchanged between the leptons and the nucleus by $q_1$.%
\fn{
Here and in what follows, the formulation `the pair neutrino-antineutrino' (or simply `the neutrinos') represents the final state $\nu_{\mu}$ and $\bar{\nu}_e$. For our purposes, the two particles can be treated together. In the end, indeed, the phase space of the two particles can be simplified to that of a single particle (for details, cf. ref.~\cite{Pruna:2019CHIPP}).
}
In the infinitely heavy nucleus limit, 3-momentum is not conserved in the photon-nucleus vertex (the nucleus acts as a sink for 3-momentum).
Choosing for convenience the electron momentum to lie along the $z$ axis, the explicit parametrization of the four-momenta is 
\ali{
p &= \Big(\sqrt{m_{\mu}^2 + |\vec{p}|^2}, \vec{p}\Big), 
&
p' &= \Big(E_e,0,0,-\sqrt{E_e^2 - m_e^2 }\Big), \nonumber \\
q_1 &=(q_{1_0},\vec{q}_1),
&
q_2 &= (q_{2_0},\vec{q}_2).
% k = (M_N,\vec{0}),
% \qquad
% k' = (\sqrt{M_N^2 + |\vec{k'}|^2},\vec{k'}).
}
For the purpose of power counting, the muon is assumed to be non-relativistic, with $|\vec{p}| \sim m_{\mu} Z \alpha$. Since we focus on the electron endpoint, i.e. $E_e \simeq m_\mu$, the momentum carried by neutrinos is small $|\vec{q}_2| \simeq q_{2_0} \simeq \Delta E $. The energy transfer to the nucleus is also small, $q_{1_0} \simeq \frac{m_\mu^2}{M_N} \simeq \Delta E$, but the spatial momentum is large, $|\vec{q}_1| \sim m_{\mu}$.

Aiming at the $\mathcal{O}(\alpha)$ NLO (hard) matching between EFTs II and III, it is sufficient to parameterize the scattering amplitude with three form-factors $F_i^{\II}$ ($i =a,b,c$), which are in general a function of the momentum transfer $q_1^2 =(p-p'-q_2)^2$. We thus write the scattering amplitude as
\ali{
i \mathcal{A}^{\II} = 16 \pi \, Z \alpha \, \sqrt{2} G_F \bar{u}_e \bigg\{
F_a^{\II} \dfrac{\gamma_{\rho}}{m_{\mu}^3} P_L 
+
F_{b}^{\II} \frac{p_{\rho}}{m_{\mu}^4} P_R 
+
F_{c}^{\II} \dfrac{p'_{\rho}}{m_{\mu}^4} P_R \bigg\} u_{\mu} \bar{u}_{h_N} u_{h_N} \bar{u}_{\nu_{\mu}} \gamma^{\rho} P_L v_{\nu_e}.
}
These form-factors, which correspond to the hard region of the loop momenta of the SC approach, are directly related to the (hard) matching coefficients of EFT III. Due to their length, explicit expressions are omitted here.

\subsection{EFT III: \texorpdfstring{$\mu \sim \mu_{\shvar}$}{mu ≃ mus}}

The LP matching conditions between EFTs II and III relate the scattering amplitude discussed before to the matrix element of the operators defined at the semi-hard scale,
\ali{
\label{eq:matching-II-III}
\mathcal{A}^{\II} &=  \sum_{i=1}^{3}\Big\langle e N \bar{\nu}_{e} \nu_{\mu} \Big| \int dt \bigg\{ C_{i}^{\III}\ar{t} \mathcal{O}_{i}^{\III}\ar{t} \bigg\} \Big| \mu N \Big\rangle.
}
Here, $\mathcal{O}_{i}^{\III}$ are the EFT III operators, given by
\ali{
\mathcal{O}_{i}^{\III}\ar{t} &\equiv \bar{h}_N^{\sh}\ar{0} h_N^{\sh}\ar{0} \left[\bar{\xi}^{\hc} W^{\hc}\right]\ar{t n_+} \, \Gamma_{i}^{\ell}\, Y_{n_-}^{\sh\dagger}\ar{0} \psi^{\p}\ar{0} \bar{\nu}_{\mu}^{\sh}\ar{0} \, \Gamma_{i}^{\nu} \, \nu_{e}^{\sh}\ar{0},
}
where $\xi^{\hc}$ represents a hard-collinear electron field and $\psi^{\p}$ the potential muon field, and where $W^{\hc}$ and $Y_{n_-}^{\sh}$ are hard-collinear and semi-hard Wilson lines, respectively (for details, cf. ref.~\cite{Fontes:2024yvw}). Finally, the following Dirac structures appear in the SM:
\bs
\label{eq:Gammas}
\begin{align}
\hspace{20mm}
\Gamma_{1}^{\ell} &= \frac{\slashed n_+}{2}  P_L,
&
\Gamma_{1}^{\nu} &= \slashed n_- P_L, \hspace{20mm} \\
\Gamma_{2}^{\ell} &= \gamma_\perp^\mu P_L,
&
\Gamma_{2}^{\nu} &= \gamma_{\perp \mu}
P_L,\\
\Gamma_{3}^{\ell} &= \frac{\slashed n_+}{2}  P_L,
&
\Gamma_{3}^{\nu} &= \slashed n_+ P_L.
\end{align}
\es
We now define, for convenience, the normalized matching coefficient $\tilde{C}_{i}^{\III}(\mu)$:
\ali{
\label{eq:def-for-conv}
C_{i}^{\III}(\mu) \equiv \dfrac{16 \pi \, Z \alpha(\mu) \, \sqrt{2} G_F}{m_{\mu}^3} \tilde{C}_{i}^{\III}(\mu).
}
Although, in principle, the matching should be performed for arbitrary electron energy $E_e \sim m_\mu$, we assume from the very beginning that $E_e = m_{\mu}$. This choice is justified both by the considerable simplification it brings to the expressions, as well as by the circumstance that the resulting factorization theorem will ultimately depend only on the hard matching coefficients evaluated at this energy.  Accordingly, up to $\mathcal{O}(\alpha)$, the coefficients $\tilde{C}_{i}^{\III}$ read
\bs
\label{eq:hard-functions}
\ali{
\label{eq:hard-function-1}
& \tilde{C}^{\III}_1(2 m_{\mu}, m_\mu;\mu_{\hvar}) = 1 - \frac{\alpha(\mu_{\hvar})}{4 \pi } \Bigg\{2
\ln \left(\frac{4 m_{\mu}}{\mu_{\hvar}}\right) \ln \left(\frac{m_{\mu}}{\mu_{\hvar}}\right) - \frac{23}{3} \ln \left(\frac{m_{\mu}}{\mu_{\hvar}}\right) + 4.36248 \Bigg\}, \\
\label{eq:C3III}
& \tilde{C}^{\III}_2(2 m_{\mu}, m_\mu;\mu_{\hvar}) = -1.39732 \, \dfrac{\alpha(\mu_{\hvar})}{4 \pi},
 \\
& \tilde{C}^{\III}_3(2 m_{\mu}, m_\mu;\mu_{\hvar}) = -0.24756 \, \dfrac{\alpha(\mu_{\hvar})}{4 \pi}.
}
\es
As the matching is performed at the hard scale, we use the on-shell subtraction scheme in EFT II to fix both the muon mass counterterm and the muon contribution to the $\alpha$ renormalization (i.e. to the vacuum polarization counterterm). 
The electron contribution is renormalized in the $\overline{\rm MS}$ scheme. This mixed scheme ensures that the muon mass does not run below the hard scale, and that $\alpha$ only runs with one flavor (the electron one). Furthermore, it allows a straightforward comparison with the results of~\cite{Szafron:2015kja}. 

To obtain the matching coefficients at arbitrary scales, we define $U_{\hvar,i}$ as the renormalization group equation (RGE) evolution factor. It is obtained by solving the RGE for the coefficients $C_{i}^{\III}$, with the initial conditions at the hard scale. 
The fact that $\tilde{C}_{2}^{\III} \sim \tilde{C}_{3}^{\III} \sim \mathcal{O}(\alpha)$ implies that $C_{2}^{\III}$ and $C_{3}^{\III}$ are not relevant at leading logarithmic (LL) and next-to-leading logarithmic (NLL) accuracy.
The RGE for $C_{1}^{\III}$ and its solution $U_{\hvar,1}$ are the same as those for $C_{X}^{\III}$ of ref.~\cite{Fontes:2024yvw}. This identity holds since
a) the neutrinos do not modify the running,
b) the identity $\bar{\xi}^{\hc}  \frac{\slashed n_+}{2}  P_L  \psi^{\p} = \bar{\xi}^{\hc} P_R \psi^{\p}$ holds and
c) the operators with different Dirac structures do not mix under RGE up to NLL accuracy.
%
%This RGE is consistent with eq.~(\ref{eq:hard-function-1}) when taking into account the definition~(\ref{eq:def-for-conv}) and the fact that, in the prefactor extracted there, only $\alpha$ evolves below the hard scale.

Just as in ref.~\cite{Fontes:2024yvw}, the decoupling of the electron soft modes is required at this stage. This is achieved via the definition
\ali{
\label{soft-decoupling-xi}
\bar{\xi}^{\hc}\ar{x} &= \bar{\xi}^{\hc}_{\zz}\ar{x} Y_{n_-}^{\s\dagger}\ar{x_-},
}
and the operators $\mathcal{O}_{1-3}^{\III}$ are now written in terms of the soft-decoupled field $\bar{\xi}^{\hc}_{\zz}$.

Finally, we follow ref.~\cite{Fontes:2024yvw} in neglecting the semi-hard Wilson lines in what follows. Given our assumption $\Delta E \sim m_e$, they will not contribute to the final rate.

\subsection{EFT IV: \texorpdfstring{$\mu \sim \mu_{\svar}$}{mu ≃ mus}}

The LP matching conditions for currents between EFTs III and IV are trivial, $C_i^{\IV}( 2m_\mu,$ $ m_\mu;\mu_{\shvar}) = C_i^{\III}( 2m_\mu, m_\mu;\mu_{\shvar})$.
The most interesting features in this passage happen at the level of the Lagrangian.  The operators are now 
\label{eq:operators-IV}
\ali{
\mathcal{O}_{i}^{\IV}\ar{t} &\equiv \bar{h}_N^{\s}\ar{0} h_N^{\s}\ar{0} \left[\bar{\xi}^{\c} W^{\c}\right]\ar{t n_+} \, \Gamma_{i}^{\ell}\, Y_{n_-}^{\s\dagger}\ar{0} \Psi^{\p}\ar{0} \bar{\nu}_{\mu}^{\s}\ar{0} \, \Gamma_{i}^{\nu} \, \nu_{e}^{\s}\ar{0},
}
where the soft Wilson lines, arising from the field redefinitions in eq.~(\ref{soft-decoupling-xi})  factor out soft interactions at this stage. We perform the soft decoupling of the remaining fields via
\cite{Fontes:2024yvw}
\bs
\label{eq:soft-decoupling}
\ali{
\bar{h}_{N}^{\s} &= \bar{h}^{\s}_{N\zz} Y_{v}^{\s\dagger} , 
&
\bar{\Psi}^{\p} &= \Psi^{\p}_{\zz}  Y_{v}^{\s\dagger},
&
\bar{e}^{\s} &= \bar{e}^{\s}_{\zz}  Y_{v}^{\s\dagger}, \\
h_{N}^{\s} &= \overline{Y}_{v}^{\s} h_{N\zz}^{\s},
&
\Psi^{\p} &= \overline{Y}_{v}^{\s} \Psi^{\p}_{\zz},
&
e^{\s} &= \overline{Y}_{v}^{\s} e_{\zz}^{\s}.
}
\es
As a consequence, the operators of eqs.~(\ref{eq:operators-IV}) can be rewritten as a product 
\ali{
\label{eq:to-analogy}
\mathcal{O}_{i}^{\IV}\ar{t} &= \mathcal{Q}_{\svar,i}\ar{0} \, \mathcal{O}_{i\zz}^{\IV}\ar{t},
}
with $\mathcal{O}_{i\zz}^{\IV}$ being the soft-decoupled operators,
\label{eq:operators-IV-decoupled}
\ali{
\mathcal{O}_{i\zz}^{\IV}\ar{t} = \bar{h}_{N\zz}^{\s}\ar{0} h_{N\zz}^{\s}\ar{0} \left[\bar{\xi}^{\c} W^{\c}\right]\ar{t n_+} \Gamma_{i}^{\ell} \, \Psi_{\zz}^{\p}\ar{0},
}
and $\mathcal{Q}_{\svar,i}$ collects the \textit{soft operators}, 
\ali{
\label{eq:overall-soft-operators}
\mathcal{Q}_{\svar,i}\ar{x} = \mathcal{N}_{\svar,i} \ar{x} \, \mathcal{O}_{\svar}\ar{x},
}
with%
\fn{In ref.~\cite{Fontes:2024yvw}, there were no neutrinos, so the soft operator was simply given by $\mathcal{O}_{\svar}$.}
\ali{
\label{eq:soft-operator}
\mathcal{O}_{\svar}\ar{x} = \left[Y_{v}^{\s\dagger} \overline{Y}_{v}^{\s} \, Y_{n_-}^{\s\dagger} \overline{Y}_{v}^{\s} \right]\ar{x},
\qquad
\mathcal{N}_{\svar,i} \ar{x} &= \bar{\nu}_{\mu}^{\s}\ar{x} \, \Gamma_{i}^{\nu} \, \nu_{e}^{\s}\ar{x}.
}
Eq.~(\ref{eq:overall-soft-operators}) reveals that the soft operators $\mathcal{Q}_{\svar,j}$ are further factorized into two operators --- the neutrino operators $\mathcal{N}_{\svar,i}$ and the photon operator $\mathcal{O}_{\svar}$ --- as the neutrinos, being charge neutral, do not interact with photons. Hence, the LO matrix element of $\mathcal{N}_{\svar,j}$ is tree-level exact in QED and can be treated as a universal factor.

\subsection{EFT V: \texorpdfstring{$\mu \sim \mu_{\scvar}$}{mu ≃ mus}}

The transition to EFT V is obtained by matching the collinear electron field onto the bHQET field, which involves integrating out collinear modes. Near the DIO endpoint, the electron energy can fluctuate only by a small amount of order $\Delta E$, justifying this procedure. This matching introduces the coefficient $C_m(m_e;\mu_{\svar})$, which beyond one loop contains rapidity divergences. These cancel against contributions from the soft matrix element, which receives soft massive fermion corrections starting at two loops, leading to rapidity renormalization group equations. Since these effects lie beyond the accuracy considered in this paper, we will ignore this technical complication.%
\fn{They appear at the two-loop order \cite{Hoang:2015iva}.}

The LP matching conditions between EFTs IV and V are 
\ali{
\label{eq:matching-IV-V}
\mathcal{Q}_{\svar,i}\ar{0}  \int dt \, C_{i}^{\IV}\ar{t} \mathcal{O}_{i\zz}^{\IV}\ar{t} &= \mathcal{Q}_{\svar,i}\ar{0} 
\int dt \, C_{i}^{\IV}\ar{t}e^{i m_e  n_+ \cdot v_e t } C_m(m_e;\mu_{\svar}) \mathcal{O}_{i}^{\V}\ar{0} \;,
}
with the four-fermion operator
\ali{
\mathcal{O}_{i}^{\V}\ar{t} &\equiv \bar{h}_{N\zz}^{\s} h_{N\zz}^{\s} \, \bar{h}_e^{\sc} \, \Gamma_{i}^{\ell}\, \Psi_{\zz}^{\p}.
}
We note that, due to eq.~(\ref{eq:Gammas}),  $\mathcal{O}_{1}^{\V}=\mathcal{O}_{3}^{\V}$. The coefficient $C_m$ belongs to a class of radiative jet functions that, in QED with a massive electron, begin contributing already at LP. This contrasts with the case of QCD with massless quarks, where \textit{radiative} jet functions are purely power-suppressed objects, requiring explicit soft radiation to contribute (hence their name). 
The same function $C_m$ multiplies all operators $\mathcal{O}_{i}^{\V}$ in eq.~(\ref{eq:matching-IV-V}) and it also appears in muon conversion \cite{Fontes:2024yvw}. 
This illustrates the universality of the low energy matrix elements defined within the modern EFT approach. 

For convenience, we define the complete matching coefficient in the EFT V as
\ali{
\label{eq:CV}
C_{i}^{\V}(\mu_{\svar}) \equiv  C_{i}^{\IV}(2 m_{\mu},m_\mu;\mu_s) C_m(m_e; \mu_{\svar}) = \int dt e^{i m_e n_+ \cdot v_e t} C_i^{\IV}\ar{t} C_m(m_e; \mu_{\svar}).
}
Finally, we perform the soft-collinear decoupling via the definitions \cite{Fontes:2024yvw}
\bs
\label{eq:sc-decoupling}
\ali{
\bar{h}_{N\zz}^{\s} &= \bar{h}^{\s}_{N\zzzz} Y_{n_+}^{\sc\dagger}, 
&
\bar{\Psi}_{\zz}^{\p} &= \bar{\Psi}^{\p}_{\zzzz} Y_{n_+}^{\sc\dagger}, 
&
\bar{h}_{e}^{\sc} &= \bar{h}^{\s}_{e\zz} Y_{v_e}^{\sc\dagger}, \\
h_{N\zz}^{\s} &= \overline{Y}_{n_+}^{\sc} h_{N\zzzz}^{\s},
&
\Psi_{\zz}^{\p} &= \overline{Y}_{n_+}^{\sc} \Psi^{\p}_{\zzzz},
&
h_{e}^{\sc} &= \overline{Y}_{v_e}^{\sc} h_{e\zz}^{\sc}.
}
\es
This leads to a factorization analogous to that of eq.~(\ref{eq:to-analogy}),
\ali{
\mathcal{O}_{i}^{\V} &= \mathcal{O}_{\scvar}\ar{0} \, \mathcal{O}_{i\zz}^{\V},
}
with $\mathcal{O}_{\scvar}$ being the soft-collinear operator \cite{Fontes:2024yvw}, composed of the soft-collinear Wilson lines,
\ali{
\mathcal{O}_{\scvar}\ar{x} = 
\left[Y^{\sc\dagger}_{n_+} \overline{Y}_{n_+}^{\sc} Y^{\sc\dagger}_{v_e} \overline{Y}_{n_+}^{\sc}\right]\ar{x},
}
and $\mathcal{O}_{i\zz}^{\V}$ the decoupled four-fermion operators,
\ali{
\mathcal{O}_{i\zz}^{\V}\ar{t} &\equiv \bar{h}_{N\zzzz}^{\s} h_{N\zzzz}^{\s} \, \bar{h}_{e\zz}^{\sc} \, \Gamma_{i}^{\ell}\, \Psi_{\zzzz}^{\p}.
}

\section{Factorization}
\label{sec:Factorization}

\subsection{Factorization theorem}
\label{subsec:Factorization-theorem}

Having derived the operators at the soft-collinear scale, we are ready to consider the physical observable --- the differential decay rate for muon eDIO, for which we derive a factorization theorem. 
The rate can be calculated using standard methods; in particular, it is determined by the amplitude for muon DIO, in the presence of arbitrary radiation in the final state, $\mathcal{X}$. The amplitude, in turn, is obtained by inserting the current between the initial and final states,
\ali{
\mathcal{M}_{\mu_{H} \to e N \nu_{\mu} \bar{\nu}_e  \mathcal{X}} = \Big\langle e^{\sc} N^{\s} \bar{\nu}_{e}^{\s} \nu_{\mu}^{\s} \mathcal{X}^{\s}
\mathcal{X}^{\sc} \Big| \mathcal{J}\ar{0} \Big| \mu_H \Big\rangle,
}
where the radiation $\mathcal{X}$ is now split into a soft component ($\mathcal{X}^{\s}$) and soft-collinear ($\mathcal{X}^{\sc}$) one. 
The current $\mathcal{J}$ is defined in terms of the soft and soft-collinear operators,
\ali{
\mathcal{J}\ar{0} \equiv \sum_{i=1}^{3} C_{i}^{\V}(\mu_{\svar}) \mathcal{O}_{\scvar}\ar{0} \, \mathcal{Q}_{\svar,i}\ar{0} \mathcal{O}_{i\zz}^{\V}\ar{0},
}
where the coefficients $C_{i}^{\V}(\mu_{\svar})$, introduced in eq.~(\ref{eq:CV}), can be related to the coefficients $C_{i}^{\III}(2 m_{\mu}, m_\mu;\mu_{\hvar})$ through RGE; that is, up to NLL accuracy,
\ali{
C_{i}^{\V}(\mu_{\svar}) &= C_{i}^{\III}(2 m_{\mu}, m_\mu;\mu_{\hvar}) \, U_{\hvar,i}(\mu_{\hvar},\mu_{\svar}) \, C_m(m_e;\mu_{\svar}) \nonumber \\
&= \dfrac{16 \pi \, Z \alpha(\mu_{\svar}) \, \sqrt{2} G_F}{m_{\mu}^3} \hat{C}_{i} \, C_m(m_e;\mu_{\svar}), 
}
with $\hat{C}_{i}$ representing the hard contribution evolved to the soft scale,
\ali{
\label{eq:hatted-C-def}
\hat{C}_{i} &\equiv \dfrac{\alpha(\mu_{\hvar})}{\alpha(\mu_{\svar})} \tilde{C}_{i}^{\III}(2 m_{\mu}, m_\mu;\mu_{\hvar}) \, U_{\hvar,i}(\mu_{\hvar},\mu_{\svar}). 
}
This allows us to rewrite the current as
\ali{
\mathcal{J}\ar{0} &= \dfrac{16 \pi \, Z \alpha(\mu_{\svar}) \, \sqrt{2} G_F}{m_{\mu}^3} \sum_{i=1}^{3} \hat{C}_{i} \, C_m(m_e;\mu_{\svar}) \mathcal{O}_{\scvar}\ar{0} \, \mathcal{Q}_{\svar,i}\ar{0} \mathcal{O}_{i\zz}^{\V}\ar{0},
}
so that the amplitude becomes
\ali{
& \mathcal{M}_{\mu_{H} \to e N \nu_{\mu} \bar{\nu}_e  \mathcal{X}} = \nonumber \\
=& \dfrac{16 \pi \, Z \alpha(\mu_{\svar}) \, \sqrt{2} G_F}{m_{\mu}^3} \sum_{i=1}^{3} \hat{C}_{i}^{\III} \, C_m(m_e;\mu_{\svar}) \, 
\Big\langle \mathcal{X}^{\sc} e^{\sc} \Big| \mathcal{O}_{\scvar}\ar{0} \left[\bar{h}_{e\zz}^{\sc}\ar{0} \right]_\alpha \Big| 0 \Big\rangle  \nonumber \\
& \hspace{2mm} \times \Big\langle N^{\s} \mathcal{X}^{\s} \bar{\nu}_{e}^{\s} \nu_{\mu}^{\s} \Big|\bar{h}_{N\zzzz}^{\s}\ar{0} h_{N\zzzz}^{\s}\ar{0} \left[\Gamma_{i}^{\ell} \Psi_{\zzzz}^{\p}\ar{0}\right]_{\alpha} \mathcal{Q}_{\svar,i}\ar{0} \Big|\mu_H \Big\rangle.
}
Fields associated with each scale are grouped and treated collectively (i.e. soft and soft-collinear modes are each handled collectively), enabling the separation of soft and soft-collinear contributions.
However, the matrix elements of the sterile fields can be treated individually. For example, the soft-collinear electron field is sterile after soft-collinear decoupling, which allows us to write
\ali{
\Big\langle \mathcal{X}^{\sc} e^{\sc} \Big| \mathcal{O}_{\scvar}\ar{0} \bar{h}_{e\zz}^{\sc}\ar{0} \Big| 0 \Big\rangle
&=
\Big\langle \mathcal{X}^{\sc} \Big| \mathcal{O}_{\scvar}\ar{0} \Big| 0 \Big\rangle
\Big\langle e^{\sc} \Big| \Big[ \bar{h}_{e\zz}^{\sc}\ar{0}\Big]_{\alpha} \Big| 0 \Big\rangle \nonumber \\
&= \Big\langle \mathcal{X}^{\sc} \Big| \mathcal{O}_{\scvar}\ar{0} \Big| 0 \Big\rangle \left[\bar{u}_{h_e} \right]_\alpha.
}
In the same way, and since the potential muon defines the bound muon wave-function together with nucleus \cite{Fontes:2024yvw}, we have 
\ali{
\label{eqs:factorization-aux}
& \Big\langle N^{\s} \mathcal{X}^{\s} \bar{\nu}_{e}^{\s} \nu_{\mu}^{\s} \Big|\bar{h}_{N\zzzz}^{\s}\ar{0} h_{N\zzzz}^{\s}\ar{0} \left[\Gamma_{i}^{\ell} \Psi_{\zzzz}^{\p}\ar{0}\right]_{\alpha} \mathcal{Q}_{\svar,i}\ar{0} \Big|\mu_H \Big\rangle \nonumber \\
=& \Big\langle \mathcal{X}^{\s} \bar{\nu}_{e}^{\s} \nu_{\mu}^{\s} \Big|\mathcal{Q}_{\svar,i}\ar{0} \Big|0 \Big\rangle 
\Big\langle N^{\s} \Big| \bar{h}_{N\zzzz}^{\s}\ar{0} h_{N\zzzz}^{\s}\ar{0} \left[\Gamma_{i}^{\ell} \, \Psi_{\zzzz}^{\p}\ar{0}\right]_{\alpha} \Big| \mu_H \Big\rangle \nonumber \\
=& \Big\langle \mathcal{X}^{\s} \bar{\nu}_{e}^{\s} \nu_{\mu}^{\s} \Big|\mathcal{Q}_{\svar,i}\ar{0} \Big|0 \Big\rangle  \dfrac{1}{\sqrt{2 m_{\mu}}} \bar{u}_{h_N} u_{h_N} \, \left[\Gamma_{i}^{\ell} \, u_{\Psi}\right]_\alpha \psi_{\rm Schr.}\ar{0}
,
}
where $\psi_{\mathrm{Schr.}}\ar{x}$ denotes the position-space wave function of the muon in the $1s$ state of a hydrogen-like ion. Then, using eq.~(\ref{eq:overall-soft-operators}), we rewrite the amplitude as
\ali{
\mathcal{M}_{\mu_{H} \to e N \nu_{\mu} \bar{\nu}_e  \mathcal{X}} &= \dfrac{16 \pi \, Z \alpha(\mu_{\svar}) \, \sqrt{2} G_F}{m_{\mu}^3} \, \dfrac{\psi_{\rm Schr.}\ar{0}}{\sqrt{2 m_{\mu}}} \sum_{i=1}^{3}
\bar{u}_{h_N} u_{h_N} \, \bar{u}_{h_e}  \Gamma_{i}^{\ell} \, u_{\Psi}  \,  \hat{C}_{i} \, C_m(m_e;\mu_{\svar}) \nonumber \\
& \hspace{-17mm} \times 
\big\langle \mathcal{X}^{\sc} \big| \mathcal{O}_{\scvar}\ar{0} \big| 0 \big\rangle
\,\,
\big\langle \mathcal{X}^{\s} \big| \mathcal{O}_{\svar}\ar{0} \big| 0 \big\rangle 
\,\, \big\langle \bar{\nu}_{e}^{\s} \nu_{\mu}^{\s} \big| \mathcal{N}_{\svar,i} \ar{0} \big| 0\big\rangle.
}
This allows us to finally consider the decay rate. Following ref.~\cite{Fontes:2024yvw}, we write it as
\ali{
\label{eq:decay-width}
\Gamma_{\mu_{H} \to e N \nu_{\mu} \bar{\nu}_e \mathcal{X}}
&= \dfrac{1}{2 M_{\mu_{H}}} \int (2 \pi)^4 \delta^{(d)} \Big(p_{\mu_{H}}-p'-k' - p_{\nu} - p_{\bar{\nu}} - \sum_{i}p_{\mathcal{X}i}\Big) \dfrac{d^{3} k'}{(2 \pi)^{3} 2 M_N}  \nonumber \\
& \hspace{1mm} \times \dfrac{d^{3} p'}{(2 \pi)^{3} 2 E_e} \dfrac{d^{3} p_{\nu}}{(2 \pi)^{3} 2 E_{\nu}} \dfrac{d^{3} p_{\bar{\nu}}}{(2 \pi)^{3} 2 E_{\bar{\nu}}}  d\mathcal{P}^{\s} \, d\mathcal{P}^{\sc} \,\, \big|\overline{\mathcal{M}}_{\mu_{H} \to e N \nu_{\mu} \bar{\nu}_e \mathcal{X}_{e} \mathcal{X}}\big|^2,
}
with $d\mathcal{P}^{\s}$ and $d\mathcal{P}^{\sc}$ being the phase space factors of the emitted real radiation~\cite{Fontes:2024yvw}. All the integrals can be performed in $4$ dimensions; however, to maintain consistency with the scheme defined by the SC approach, we keep $d\mathcal{P}^{\s}$ and $d\mathcal{P}^{\sc}$ in $d$ dimensions. 
We now define the soft function and the soft-collinear function respectively as~\cite{Fontes:2024yvw} 
\bs
\label{eq:S-and-SC-functions}
\begin{align}
\label{eq:S-function}
\mathcal{S}(E_{\svar}) &\equiv \sum_{\mathcal{X}^{\s}}\int \prod_{i} \frac{d^{d-1} p_{\mathcal{X}_i^{\s}}}{(2\pi)^{d-1}2E_{\mathcal{X}_i^{\s}}} 
\delta(E_{\svar} - E_{\mathcal{X}^{\s}}) \langle 0 |\mathcal{O}^\dagger_{\svar}\ar{0} | \mathcal{X}^{\s} \rangle  \langle \mathcal{X}^{\s}|\mathcal{O}_{\svar}\ar{0} \left| 0 \right\rangle, \\
% \nonumber \\
% %
% &= \int \frac{dt}{2\pi} e^{i t E_{\svar}}\langle 0 |\mathcal{O}_{\svar}\ar{t}  \mathcal{O}_{\svar}\ar{0} \left| 0 \right\rangle, \\
%
\label{eq:SC-function}
\mathcal{SC}(E_{\scvar}) &\equiv \sum_{\mathcal{X}^{\sc}}\int \prod_{i} \frac{d^{d-1} p_{\mathcal{X}_i^{\sc}}}{(2\pi)^{d-1}2E_{\mathcal{X}_i^{\sc}}}  \delta(E_{\scvar} - E_{\mathcal{X}^{\sc}}) \langle 0 |\mathcal{O}^\dagger_{\scvar}\ar{0} | \mathcal{X}^{\sc} \rangle  \langle \mathcal{X}^{\sc}|\mathcal{O}_{\scvar}\ar{0} \left| 0 \right\rangle.
% \nonumber \\
% &= \int \frac{dt}{2\pi} e^{i t E_{\scvar}}\langle 0 |\mathcal{O}_{\scvar}\ar{t}  \mathcal{O}_{\scvar}\ar{0} \left| 0 \right\rangle.
\end{align}
\es
After calculating the neutrino matrix element and evaluating the phase-space integrals, we arrive at the final form of the factorization theorem:
\ali{
\label{eq:factorized}
\Gamma'_{\mu_{H} \to e N \nu_{\mu} \bar{\nu}_e \mathcal{X}} &= \dfrac{1024 \, \Gamma_0 \, Z^5 \alpha(\mu_{\svar})^5}{5 \pi m_{\mu}^6} |\psi_{\rm corr}|^2 \, \Bigg[\big|\hat{C}_{1}\big|^2 + 2 \big|\hat{C}_{2}\big|^2 + \big|\hat{C}_{3}\big|^2 \Bigg] \left|C_m(m_e;\mu_{\svar})\right|^2 \nonumber \\
& \hspace{-10mm} \times \int_{-\infty}^{\infty} d E_{\scvar} \int_{-\infty}^{\infty} d E_{\svar} \mathcal{S}(E_{\svar}) \, \mathcal{SC}(E_{\scvar}) \,\, (\Delta E - E_{\svar} - E_{\scvar})^{5} \,\, \theta(\Delta E - E_{\svar} - E_{\scvar}),
}
where $|\psi_{\rm corr}|^2$ denotes  the correction to the Schrödinger wavefunction of the bound muon~\cite{Szafron:2015kja}.

\subsection{Formul\ae}
\label{sec:we-love-latin}

Inserting LO expressions for all the component functions, we trivially recover eq.~(\ref{eq:rate-LO}). 
Expanding all the terms to NLO (without RGE effects) and using $\tilde{C}_1=1 + \mathcal{O}(\alpha)$, we find the fixed order NLO expression
\ali{
\label{eq:factorized-NLO}
\Gamma^{'\textrm{NLO}}_{\mu_{H} \to e N \nu_{\mu} \bar{\nu}_{e} \mathcal{X}} &= \dfrac{1024 \, \Gamma_0 \, Z^5 \alpha(\mu_{\svar})^5 \Delta E^5}{5 \pi m_{\mu}^6} |\psi_{\rm corr}|_{\rm{NLO}}^2 \, \bigg[1 + 2 \tilde{C}_1^{\III}\Big|_{\mathcal{O}(\alpha)} \bigg] \left|C_m(m_e;\mu)\right|_{\rm{NLO,OS}}^2 \nonumber \\
& \hspace{-25mm} \times
\Bigg\{1 + \int_{-\infty}^{\Delta E} d E_{\svar} \mathcal{S}(E_{\svar})\big|_{\mathcal{O}(\alpha)} \left(\frac{\Delta E - E_{\svar}}{\Delta E}\right)^{5} 
+\int_{-\infty}^{\Delta E} d E_{\scvar} \, \mathcal{SC}(E_{\scvar})\big|_{\mathcal{O}(\alpha)} \left(\frac{\Delta E - E_{\scvar}}{\Delta E}\right)^{5}  \Bigg\},
}
where we have to consistently drop terms beyond the NLO accuracy when multiplying out the terms. 
While the $\mathcal{O}(\alpha)$ component of $\tilde{C}_1^{\III}$ can be read from eqs.~(\ref{eq:hard-functions}), we have $|\psi_{\rm corr}|_{\rm{NLO}}^2 = 1 + \frac{\alpha(\mu_{\svar})}{\pi} \times 6.4$ \cite{Fontes:2024yvw} and 
\bs
\label{eq:more-formulae}
\ali{
\label{eq:Cm}
\left|C_m(m_e;\mu)\right|_{\rm{NLO, OS}}^2 &= 1 + \frac{\alpha(\mu_{\svar})}{2 \pi} \Bigg\{ 2 \ln^2\left(\frac{m_e}{\mu} \right) - \ln \left(\frac{m_e}{\mu} \right) \df{-\frac{8}{3} \ln \left(\frac{m_e}{\mu} \right)}+ \frac{\pi^2}{12} + 2 \Bigg\}, \hspace{10mm} \\[2mm]
& \hspace{-30mm} \int_{-\infty}^{\Delta E} d E_{\svar} \mathcal{S}(E_{\svar})\big|_{\mathcal{O}(\alpha)} \left(\frac{\Delta E - E_{\svar}}{\Delta E}\right)^{5}   \nonumber \\[-1mm]
& = \dfrac{\alpha(\mu_{\svar})}{\pi} \Bigg\{ \ln^2\left(\frac{2 \Delta E}{\mu}\right)-\frac{167}{30} \ln \left(\frac{2 \Delta E}{\mu}\right) - \frac{\pi^2}{8} + \frac{17929}{1800} \Bigg\}, \\[2mm]
& \hspace{-30mm} \int_{-\infty}^{\Delta E} d E_{\scvar} \mathcal{SC}(E_{\scvar})\big|_{\mathcal{O}(\alpha)} \left(\frac{\Delta E - E_{\scvar}}{\Delta E}\right)^{5} \nonumber \\[-1mm]
& \hspace{-3mm} = \dfrac{\alpha(\mu_{\svar})}{\pi} \Bigg\{ -\ln^2\left(\frac{\Delta E m_e}{m_{\mu} \mu}\right)+\frac{107}{30} \ln\left(\frac{\Delta E m_e}{m_{\mu}\mu}\right) -\frac{\pi^2}{24}-\frac{7909}{1800} \Bigg\}.
}
\es
The coefficient $C_m(m_e;\mu)$, which is  the matching coefficient introduced in eq.~(\ref{eq:matching-IV-V}), is originally renormalized using the $\overline{\rm MS}$ subtraction scheme. To convert the final expression to the on-shell subtraction scheme, we supplemented $C_m(m_e;\mu)$ by a term arising from the $\alpha$ renormalization in on-shell subtraction, and denoted the resulting expression by $C_m(m_e;\mu)_{\rm{NLO, OS}}$. This scheme conversion term, written in blue in eq.~(\ref{eq:Cm}), was included to maintain consistency with ref.~\cite{Szafron:2015kja}, which used on-shell subtraction.%
\fn{Without this term, $C_m$ is a universal matching coefficient, identical for instance in muon conversion~\cite{Fontes:2024yvw}.}
By combining eqs.~(\ref{eq:factorized-NLO}) and (\ref{eq:more-formulae}), we confirm that eq.~(\ref{eq:rate-NLO}) is correctly reproduced. When considering the resummation, we ignore the additional scheme conversion factor and use $C_m(m_e;\mu)$ as given in ref.~\cite{Fontes:2024yvw}.

In order to resum the large logarithmic corrections, the scale of each function is set to its canonical scale (thus eliminating large logs in the matching coefficients and EFT matrix elements). The large logarithmic corrections are thus contained in the RG running factors. We choose the soft scale $\mu_{\svar}$ as the ultimate scale, to which all functions are evolved. As central values, we set the hard scale to $\mu_{\hvar} = 2m_{\mu}$ and the soft scale to $\mu_{\svar} = m_{e}$. To assess the theoretical uncertainty from scale variation, we perform a standard 7-point variation: each scale is independently varied by factors of $1/2$ and $2$ around its central value, while the two extreme combinations are omitted.
As mentioned above, $U_{\hvar,1}$ (the running of $C_{1}^{\III}$) is given in ref.~\cite{Fontes:2024yvw}, whereas the resummed expression for  $\mathcal{SC}$ function can be obtained by abelian exponentiation,
\ali{
\label{eq:resummed-SC}
  &\int \mathcal{SC}(E_{\scvar}) \left(\frac{\Delta E - E_{\scvar}}{\Delta E}\right)^{5} \,\, \theta(\Delta E - E_{\scvar}) \nonumber \\
  & \hspace{20mm} = \exp\Bigg[ \dfrac{\alpha}{\pi} \Bigg\{ -\ln^2\left(\frac{\Delta E m_e}{m_{\mu} \mu}\right)+\frac{107}{30} \ln\left(\frac{\Delta E m_e}{m_{\mu}\mu}\right) -\frac{\pi^2}{24}-\frac{7909}{1800} \Bigg\} \Bigg].  
}
We checked that, in $U_{\hvar,1}$, the running of $\alpha$ between $\mu_{\hvar}$ and $\mu_{\svar}$ is negligible (as expected due to the smallness of the QED coupling; more specifically, its induces a correction of $\mathcal{O}(0.01\%)$).%
\fn{$\alpha(\mu)$ does not run when the $ \mathcal{SC}$ function is evolved from its natural scale to the soft scale, since that natural scale is below the electron mass scale.}
Neglecting it, we obtain next-to-double logarithmic accuracy; with $\mu_{\hvar} = 2m_{\mu}$ and $\mu_{\svar} = m_{e}$, the differential rate in RG-improved perturbation theory then reads
\ali{
\label{eq:resummed}
\hspace{1mm} \Gamma^{'\textrm{NLO}}_{\mu_{H} \to e N \nu_{\mu} \bar{\nu}_e \mathcal{X}} \bigg|_{\mathrm{Resu.}} &= \dfrac{1024 \, \Gamma_0 \, Z^5 \alpha(\mu_{\svar})^5 \Delta E^5}{5 \pi m_{\mu}^6} \times 
\Big[1.023 - 0.006818\,\ln(\Delta E) \nonumber \\
& \hspace{-25mm} + 0.002403\,\ln^{2}(\Delta E) \Big] \times \mathrm{\exp} \Bigg\{\frac{\alpha}{90 \pi}  \bigg[\left(50 \frac{\alpha}{\pi} - 90\right) \ln^2 \left(\frac{2 m_{\mu}}{m_e}\right) + 225 \ln \left(\frac{2 m_{\mu}}{m_e}\right) \nonumber \\
& \hspace{15mm} -3  \ln \left(\frac{m_{\mu}}{\Delta E}\right) \left(30 \ln \left(\frac{m_{\mu}}{\Delta E}\right) + 107\right)\bigg]\Bigg\},
}
where the square-bracketed prefactor comes from $|\psi_{\rm corr}|_{\rm{NLO}}^2$, from the ratio $\alpha(\mu_{\hvar})/\alpha(\mu_{\svar})$ in eq.~(\ref{eq:hatted-C-def}) and from the small finite pieces from the different functions after they are set to their canonical scales.
We follow the procedure and conventions (of LL, NLL, etc.) from ref.~\cite{Fontes:2024yvw}.

\section{Numerical results and discussion}
\label{sec:Results}

\begin{figure}[h!]
\centering
\includegraphics[width=0.75\textwidth]{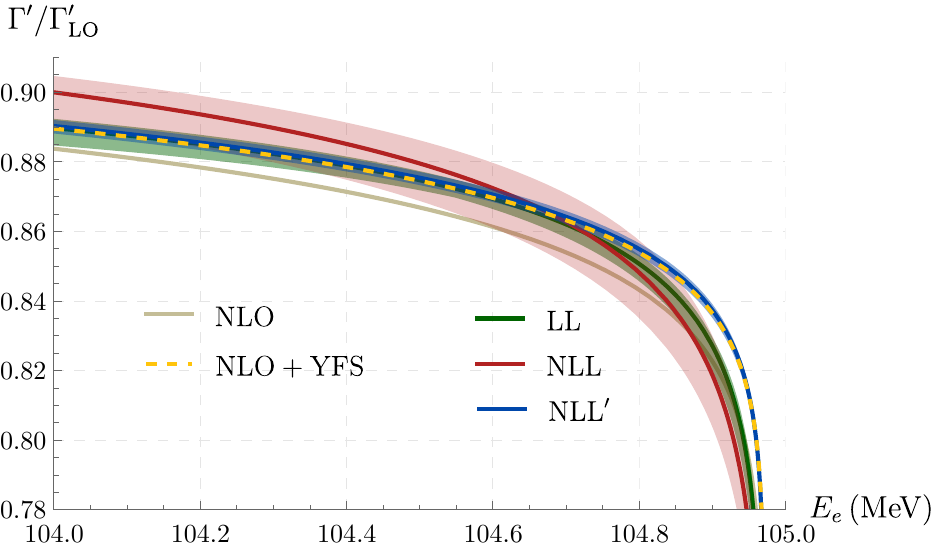}
\caption{Differential rate for eDIO against $E_e$, normalized to the LO result, for different approaches. See text for details.}
\label{fig:LO-vs-NLO}
\end{figure}

We now turn to our numerical results. We use the input values~\cite{ParticleDataGroup:2024cfk}
\ali{
m_{\mu} &= 105.658 \, \textrm{MeV}, 
&
m_e &= 0.510999 \, \textrm{MeV},\nonumber \\
\alpha(\mu_{\svar}) &= \alpha_{\textrm{OS}} = 1/137.035,
&
G_F &= 1.166378 \times 10^{-11} \, \textrm{MeV}^{-2},
}
where $\alpha_{\textrm{OS}}$ is the coupling in on-shell subtraction.%
\fn{While the on-shell subtraction scheme fixes the value of $\alpha$ in the so-called Thomson limit (i.e. for vanishing photon momentum), that value is the same as $\alpha(\mu_{\svar})$, since $\alpha$ does not run below the soft scale.}
Fig. \ref{fig:LO-vs-NLO} displays various approaches to the differential rate, normalized to the LO prediction.
The NLO and NLO+YFS curves (in solid beige and dashed yellow, respectively) correspond to those on the right panel of fig. \ref{fig:sc-LO-vs-NLO}, but are restricted to a tighter range. 
In addition, the plot includes different approaches with our EFT framework: LL (green), NLL (red) and NLL' (blue). Each of these curves is accompanied by its associated 7-point variation band, as described in section \ref{sec:we-love-latin}.
%We note that, while the  red and beige curves are obtained in the on-shell subtraction scheme, the green ones are computed in the $\overline{\rm MS}$ scheme. 

The transition from LL to NLL accuracy leads to noticeable shifts, largely driven by the inclusion of the factor $|\psi_{\rm corr}|^2$, which in the on-shell subtraction scheme corresponds to a single logarithm and is absent from the LL result. At NLL' order, the uncertainty band becomes significantly narrower than at NLL, reflecting the improved perturbative stability of the expansion. The NLL and NLL' bands overlap across the entire range of $E_e$ displayed, except in the extreme endpoint region (large $E_e$, or equivalently very small $\Delta E$), where the assumption $\Delta E \sim m_e$ breaks down and further resummation would be required. Interestingly, the NLO+YFS curve lies very close to the NLL' prediction throughout the range. It should be emphasized, however, that the NLL' result provides a consistent EFT treatment with systematic resummation of large logarithms and a well-defined estimate of theoretical uncertainties, while the NLO+YFS curve only implements a partial resummation tied to the on-shell subtraction scheme. For the present accuracy and chosen parameters, the latter nonetheless provides a very good approximation to the NLL' result.

This result --- the NLL' curve --- is the most precise prediction to date of the spectrum of eDIO, describing a correction to the LO differential result of $12.30\%$ for $\Delta E = m_e$. For phenomenological applications, we provide an expression for that curve; defining $\delta E \equiv \Delta E/E_e^{\rm max}$, we have
\ali{
\frac{\Gamma^{'\textrm{NLL'}}}{\Gamma^{'\textrm{LO}}} =
\frac{
  (\delta E)^{0.008285}
  \Big(
    0.9930
    + \bigl[0.01480 + 0.002287\,\ln(\delta E)\bigr]\,\ln(\delta E)
  \Big)}{\exp\!\Bigl(0.002323\,\bigl[\ln(\delta E) - 0.006527\bigr]^{2}\Bigr)}.
}
We also obtain the number of DIO events near the endpoint. To that end, we integrate the differential spectrum $\Gamma^{'\textrm{X}}$, where $\textrm{X}$ denotes the level of accuracy considered. The integration is performed over an interval of size $\Delta E'$ (set by the radiation cut) up to the endpoint energy $E_e^{\max}$. We define the ratio between this integral at accuracy $\textrm{X}$ and the corresponding LO integral as the $K_{\rm X}$ factor, 
\begin{align}
    K_{\rm X}(\Delta E') \equiv  \frac{\int_{E_e^{\max}-\Delta E'}^{E_e^{\max}} 
    \Gamma^{'\textrm{X}} \; d E_e}{  \int_{E_e^{\max}-\Delta E'}^{E_e^{\max}} 
    \Gamma^{'\rm LO} \; d E_e}.
\end{align}
In table \ref{table:numbers}, we show $K_{\rm X}$ for several values of the cut on the radiation around $\Delta E' \sim m_e$, and for $\rm{X}=\textrm{LL}, \textrm{NLL}, \textrm{NLL'}$. 
\begin{table}[h!]
\begin{center}

\begin{tabular}{lrrrrr}
\hlinewd{1.1pt} \\[-5mm]
$\Delta E' \, ({\rm MeV})$
& \multicolumn{1}{c}{0.25}
& \multicolumn{1}{c}{0.5}  
& \multicolumn{1}{c}{0.75}
& \multicolumn{1}{c}{1}
& \multicolumn{1}{c}{1.25}
\\[-0.5mm] \midrule
$K_{\text{LL}}(\%)$ & $85.59^{+0.26}_{-0.44}$ & $87.23^{+0.26}_{-0.30}$ & $88.10^{+0.27}_{-0.38}$ & $88.69^{+0.27}_{-0.46}$ & $89.13^{+0.26}_{-0.53}$ \\[1mm]
$K_{\text{NLL}}(\%)$ & $85.50^{+0.87}_{-1.27}$ & $87.63^{+0.69}_{-1.10}$ & $88.81^{+0.59}_{-1.00}$ & $89.61^{+0.51}_{-0.93}$ & $90.22^{+0.45}_{-0.87}$ \\[1mm]
$K_{\text{NLL'}}(\%)$ & $85.91^{+0.17}_{-0.20}$ & $87.31^{+0.18}_{-0.20}$ & $88.14^{+0.18}_{-0.20}$ & $88.73^{+0.18}_{-0.20}$ & $89.20^{+0.18}_{-0.21}$ \\[1mm]
\hlinewd{1.1pt}
\end{tabular}

\caption{$K_{\rm X}$ factor, describing the reduction in the number of background events at accuracy~X with respect to LO for a given cut on the radiation energy $\Delta E'$, shown for $\textrm{X}=\textrm{LL}, \textrm{NLL}, \textrm{NLL'}$ and for several values of $\Delta E'$ around $m_e$. Each entry is quoted as $x^{+a}_{-b}$, where $x$ denotes the central value (with scales set to their central choices), and $a$ and $b$ represent the upper and lower deviations from the corresponding 7-point variation envelope.}
\label{table:numbers}
\end{center}
\end{table}

In addition to the perturbative uncertainty of the LP term discussed earlier, three further sources of theoretical error affect the radiatively corrected spectrum. First, power-suppressed terms proportional to $\Delta E/m_\mu$ become increasingly important away from the endpoint, but they are already included in the existing numerical evaluations of the full spectrum at LO. The missing contributions therefore enter only in the radiative corrections. Second, while an all-order computation in $Z \alpha$ can be performed using the numerical method outlined at the end of page 3, this approach is currently restricted to LO in $\alpha$. The first genuine subleading Coulombic contributions, of order $\alpha \times Z \alpha$, are thus absent, and once again this omission affects only the radiative corrections. These two effects introduce an additional uncertainty that we estimate to be suppressed by a factor of order $Z \alpha$ relative to the NLL' corrections computed here over the universal LL terms, and therefore expected to remain safely below the percent level, though a more precise quantification would require a dedicated study. Third, finite nuclear size effects, although nontrivial, do not modify the spectral shape at LO but are expected to induce a shift in the overall normalization. A detailed treatment of these corrections in the presence of QED radiation will be presented in future work. 

\section{Conclusions}
\label{sec:Conclusions}

A remarkable improvement in precision is anticipated in upcoming searches for muon conversion, the process of a bound muon decaying into an electron. It is therefore of the utmost importance to derive precise predictions for both muon conversion and its only irreducible background, muon decay-in-orbit near the endpoint of the electron spectrum (eDIO). Due to bound-state effects and an intricate hierarchy of scales, however, precise calculations for both processes are remarkably complex. In a previous work, we presented a formalism to address this problem, resorting to a multiplicity of EFT techniques \cite{Fontes:2024yvw}. There, we applied the formalism to muon conversion, deriving a factorization theorem, as well as the matching coefficients and their RGEs.

In this paper, we applied the formalism to eDIO. New challenges appear in this case, due to the presence of $\nu_{\mu}$ and $\bar{\nu}_e$ in the decay products. We derived the factorization theorem, which involves functions not present in muon conversion. On the other hand, many of the functions participating in the factorization theorem of eDIO are the same as in muon conversion, which stresses the power of the EFT formalism. Our approach describes a perturbative, consistent and systematically improvable framework for the calculation of the eDIO rate. We calculated $\mathcal{O}(\alpha)$ corrections to the relevant matching coefficients, as well as their RGEs. This allowed us to provide the most precise spectrum for eDIO, thus increasing the chances of detecting muon conversion. 

This work also opens several directions on improvement. One of them concerns the inclusion of finite nuclear size effects or recoil corrections, which can be done in a systematic way using our framework. Other directions involve going beyond the leading power or the $\mathcal{O}(\alpha)$ approximation, or investigating  more energetic real radiation, i.e. considering electron energies farther away from the endpoint.  Our framework can also be adapted to related processes involving bound-state decays, such as muonic atoms in BSM scenarios.

\section*{Acknowledgements}

All Feynman diagrams were drawn with Feyngame \cite{Harlander:2020cyh, Harlander:2024qbn,Bundgen:2025utt}, and FeynCalc \cite{Mertig:1990an,Shtabovenko:2016sxi,Shtabovenko:2020gxv} was very useful. We thank Andrei Gaponenko and Pavel Murat for discussions.
D.F. thanks the Fermilab theory group for hospitality, and the Theoretical High Energy Physics area at the University of Notre Dame for their hospitality and support.
The research of D.F. was supported by the Deutsche Forschungsgemeinschaft (DFG, German Research Foundation) under grant 396021762 - TRR 257, while that of R.S. by the U.S. Department of Energy under Grant Contract No. DE-SC0012704.

\appendix

\section{Details on the SC approach}
\label{App:Full-Theory}

In this appendix, we present details of the calculation of the muon eDIO rate using the SC approach. As suggested in section \ref{sec:Basics}, this combines Feynman diagrams involving different scales, on the one hand, with an informal treatment of both approximations and bound-state effects, on the other. We start by considering the LO case in section \ref{sec:LO}, after which we discuss the NLO case in section \ref{sec:NLO}.

\subsection{LO}
\label{sec:LO}

\begin{figure}[t!]
\centering
\includegraphics[width=0.88\textwidth]{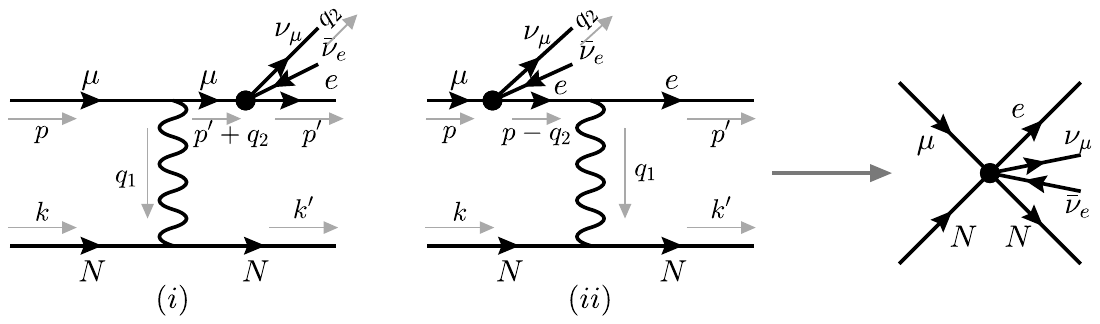}
\vspace{-135mm}
\caption{Simplification of muon DIO: in the limit of $E_e \sim m_{\mu}$ and of zero energy transfer from the leptons to the nucleus, both diagrams of figure \ref{fig:DIO-LO} reduce to a 6-fermion interaction.}
\label{fig:DIO-collapse}
\end{figure}

The bound state is a superposition of free states with different momenta, leading to
\ali{
\label{eq:bound-state-basic-amplitude}
\mathcal{M}_{\mu_{H} \to e N \nu_{\mu} \bar{\nu}_e} &= \sqrt{2 M_{\mu_H}} \int \dfrac{d^3 k}{(2 \pi)^3} \tilde{\psi}_{\rm Sch.}(\vec{k}) \, \dfrac{\mathcal{M}_{\mu N \to e N \nu_{\mu} \bar{\nu}_e}(\vec{k})}{\sqrt{2 m_{\mu}} \sqrt{2 M_N}} \nonumber \\
&\simeq \dfrac{\psi_{\rm Sch.}(0)}{\sqrt{2 m_{\mu}}} \mathcal{M}_{\mu N \to e N \nu_{\mu} \bar{\nu}_e} (\vec{k}=0),
}
where is the momentum-space equivalent of $\psi_{\mathrm{Schr.}}\ar{x}$, introduced in eq.~(\ref{eqs:factorization-aux}), and
where $\mathcal{M}_{\mu N \to e N \nu_{\mu} \bar{\nu}_e}$ is the amplitude for the free muon scattering process, 
\ali{
i \mathcal{M}_{\mu N \to e N \nu_{\mu} \bar{\nu}_e} = i \mathcal{M}_{(i)} + i \mathcal{M}_{(ii)}.
}
The Feynman diagrams corresponding to the subamplitudes $i \mathcal{M}_{(i)}$ and $i \mathcal{M}_{(ii)}$ are depicted in figure \ref{fig:DIO-collapse}.
We use the non-relativistic character of the nucleus to write, for diagram $(i)$,
\ali{
\gamma_{\alpha} u_{\mu}(p) ... \gamma_{\alpha} u_N(k) \, &\simeq \, \gamma_{0} u_{\mu}(p) ... \gamma_{0} u_N(k)
\, \simeq \,
\dfrac{\slashed{p}}{m_{\mu}} u_{\mu}(p) ... \dfrac{\slashed{k}}{M_N} u_N(k) = u_{\mu}(p) ... u_N(k).
}
A similar simplification holds for diagram $(ii)$. The kinematics of the process is approximately given by
\ali{
p&=(m_{\mu}, \vec{0}), 
&
p'&=(E_e, 0,0, E_e),
&
q_1&=(0,0,0,-E_e),
&
q_2&=(m_{\mu}-E_e,\vec{q}_2), \nonumber \\
k&=(M_N,\vec{0}),
&
k'&=(M_N,\vec{0}).
}
This leads to the following approximations:
\ali{
	& \dfrac{1}{(p-q_2)^2 - m_{e}^2} \simeq - \dfrac{1}{(p'+q_2)^2 - m_{\mu}^2} \simeq \dfrac{1}{m_{\mu}^2},
    \quad
    \gamma_0 (\slashed{p} - \slashed{q} + m_e) \simeq \dfrac{\slashed{p}}{m_{\mu}} \slashed{p} = m_{\mu},
    \nonumber \\
	& \bar{u}_e(p') \gamma_{\rho} P_L (\slashed{p}'+\slashed{q}_2 + m_{\mu}) u_{\mu}(p) \simeq \bar{u}_e(p') \left[ 2p'_{\rho} P_R + m_{\mu} \gamma_{\rho} P_L \right] u_{\mu}(p),
    \quad
    \dfrac{1}{q_1^2} \simeq - \dfrac{1}{m_{\mu}^2},
}
which in turn leads to
\ali{
    \label{eq:M-LO}
	i \mathcal{M}_{\mu N \to e N \nu_{\mu} \bar{\nu}_e} \simeq i \, 16 \pi \, Z \alpha \, \sqrt{2} G_F \mathcal{M}_{l,\rho} \mathcal{M}^{\nu,\rho} \mathcal{M}_N,
}
with
\ali{
\label{eq:Ms-LO}
\mathcal{M}^{\rho}_{\nu} = \bar{u}(p_{\nu}) \gamma^{\rho} P_L v(p_{\bar{\nu}}),
\quad
\mathcal{M}_{l,\rho} = \dfrac{p'_{\rho}}{m_{\mu}^4}\bar{u}_e(p') P_R u_{\mu}(p),
\quad
\mathcal{M}_N = \bar{u}_N(k') \, u_N(k).
}
Accordingly, both subamplitudes collapse in 6-fermion interactions, as depicted in figure \ref{fig:DIO-collapse}.
The decay width for muon DIO is then
\ali{
\label{eq:LO-decay-width}
\Gamma_{\mu_{H} \to e N \nu_{\mu} \bar{\nu}_e} &= \dfrac{1}{2 M_{\mu_H}} \int (2 \pi)^4 \delta^4(p_{\mu_H} - p' - k' - p_{\nu} - p_{\bar{\nu}}) \dfrac{d^3k'}{(2 \pi)^3 2 E_{k'}} \dfrac{d^3p'}{(2 \pi)^3 2 E_e} \dfrac{d^3p_{\nu}}{(2 \pi)^3 2 E_{\nu}} \nonumber \\
& \hspace{25mm} \times \, \dfrac{d^3p_{\bar{\nu}}}{(2 \pi)^3 2 E_{\bar{\nu}}} \left|\overline{\mathcal{M}}_{\mu_{H} \to e N \nu_{\mu} \bar{\nu}_e}\right|^2.
}
We follow ref.~\cite{Czarnecki:2011mx} in that we resort to $q_2$ to simplify the integrals of the neutrino-antineutrino pair; we find
\ali{
	\Gamma_{\mu_{H} \to e N \nu_{\mu} \bar{\nu}_e} = \int \dfrac{|\psi_{\rm Sch.}(0)|^2}{2 m_{\mu}} \dfrac{1}{2 M_{\mu_H}} & 128 \pi^2 G_F^2 Z^2 \alpha^2 dq_2^2 \nonumber \\
    & \times d\Pi_{\mu N \to e N q_2} \sum_{\rm spins} |\mathcal{M}_N|^2 |\mathcal{M}_{l,\rho\sigma}|^2 T_{\rho\sigma},
}
with
\bs
\ali{
\hspace{-2.5mm} \sum_{\rm spins} |\mathcal{M}_{l,\rho\sigma}|^2 T_{\rho\sigma} &= \dfrac{p'_{\rho} p'_{\sigma}}{m_{\mu}^8} \sum_{\rm spins} \bar{u}_{\mu}(p) P_L u_{e}(p') \bar{u}_e(p') P_R u_{\mu}(p) \, \dfrac{(-\pi)}{3 (2 \pi)^3} (q_2^2 g_{\rho\sigma} - q_{2,\rho} q_{2,\sigma}), \\
\sum_{\rm spins} |\mathcal{M}_N|^2 &= 8 M_N^2, \\
d \Pi_{\mu N \to e N q_2} &= (2 \pi)^4 \delta^4(p_{\mu_H} - p' - k' - q_2) \dfrac{d^3k'}{(2 \pi)^3 2 E_{k'}} \dfrac{d^3p'}{(2 \pi)^3 2 E_e} \dfrac{d^3q_2}{(2 \pi)^3 2 E_{q_2}}.
}
\es
In the end, we expand in powers of $\Delta E$ to find the result in eq.~(\ref{eq:rate-LO}).

\subsection{NLO}
\label{sec:NLO}

\begin{figure}[t!]
\centering
\includegraphics[width=0.491\textwidth]{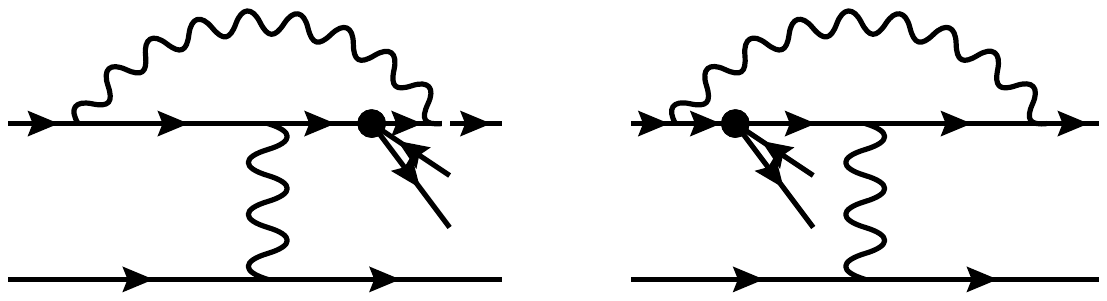}
\hspace{-5mm}
\includegraphics[width=0.49\textwidth]{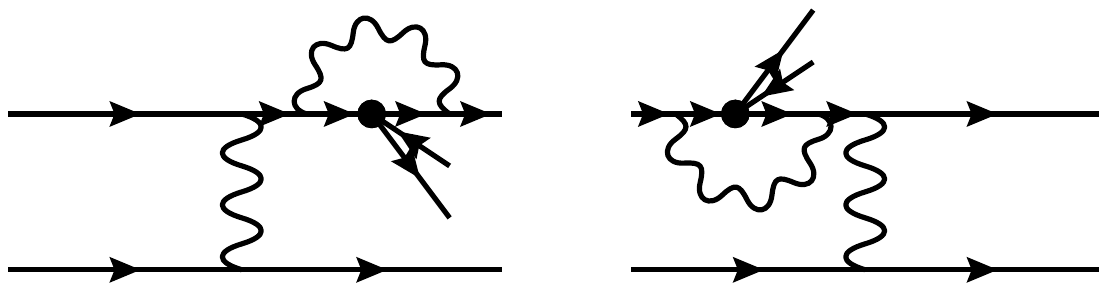} \\[-73mm]
\includegraphics[width=0.49\textwidth]{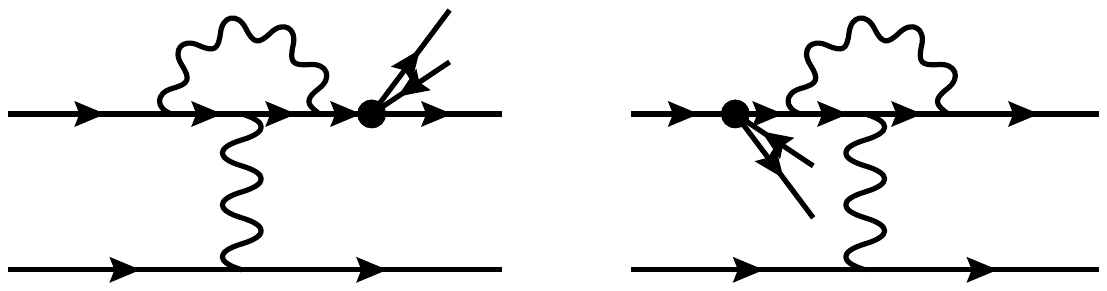}
\hspace{-5mm}
\includegraphics[width=0.49\textwidth]{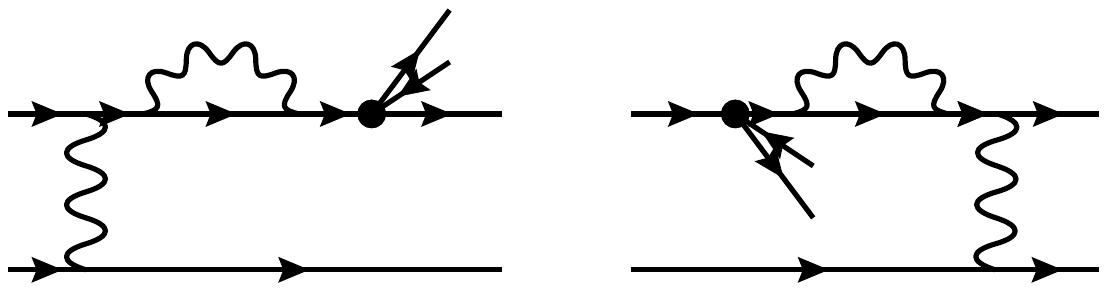} \\[-75mm]
\includegraphics[width=0.49\textwidth]{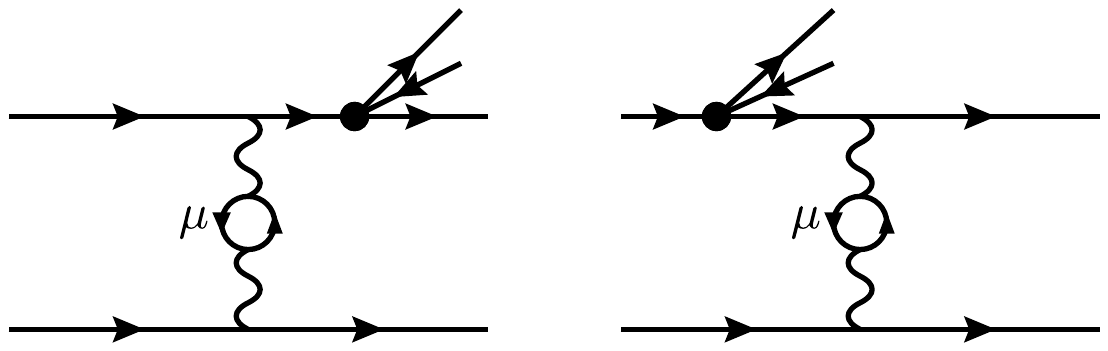}
\hspace{-5mm}
\includegraphics[width=0.49\textwidth]{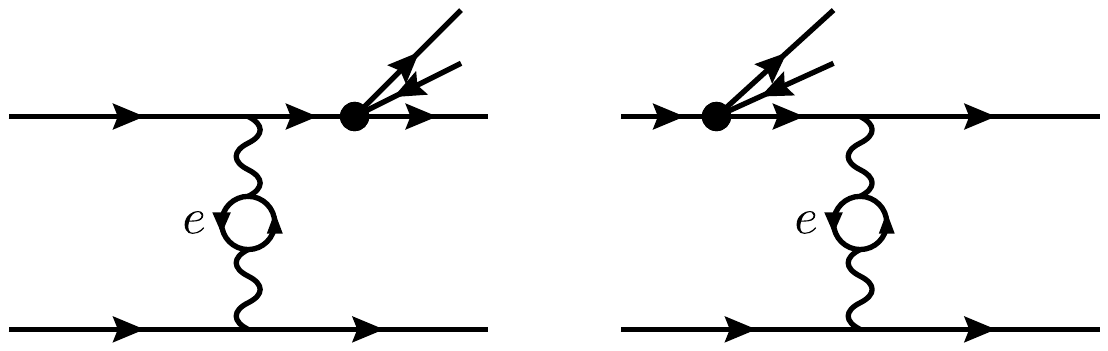}
\vspace{-71mm}
\caption{Feynman diagrams representing the NLO corrections to DIO. The particle labels are shown only whenever they are not obvious.}
\label{fig:NLO-diagrams}
\end{figure}

\begin{figure}[h!]
\hspace{-8mm}
\includegraphics[width=1.1\textwidth]{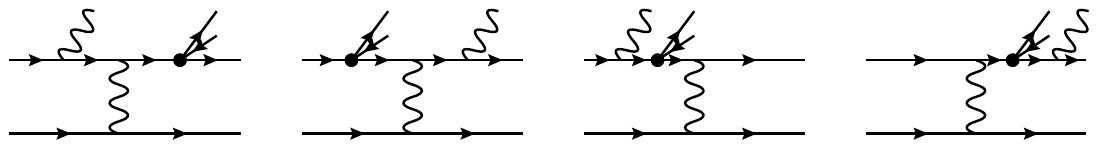}
\vspace{-192mm}
\caption{Relevant real emission diagrams in DIO in the soft approximation.}
\label{fig:DIO-real}
\end{figure}

At NLO, we consider the diagrams of figure \ref{fig:NLO-diagrams}. With the exception of the first two diagrams, each diagram has a corresponding counterterm (not displayed here), which we calculate in on-shell subtraction.
Performing approximations similar to those described in section \ref{sec:LO}, we write the NLO amplitude as a generalization of eq.~(\ref{eq:M-LO}), 
\ali{
    i \mathcal{M}^{\textrm{NLO}}_{\mu N \to e N \nu_{\mu} \bar{\nu}_e \mathcal{X}} = \mathcal{M}_{l,\rho}^{\textrm{NLO}} \mathcal{M}^{\nu,\rho} \mathcal{M}_N,
}
with $\mathcal{M}^{\nu,\rho}$ and $\mathcal{M}_N$ still given in eq.~(\ref{eq:Ms-LO}), and
\ali{
\label{eq:M-NLO}
\mathcal{M}_{l,\rho}^{\textrm{NLO}} = 16 \pi \, Z \alpha(\mu_{\svar}) \, \sqrt{2} G_F \, \, \bar{u}_e(p') \Bigg[ F_a \dfrac{\gamma^{\rho}}{m_{\mu}^3} \,  P_L + F_{b} \frac{p^{\rho}}{m_{\mu}^4} P_R + F_{c} \dfrac{p'_{\rho}}{m_{\mu}^4} P_R \Bigg] u_{\mu}(p).
}
As a consequence, the NLO amplitude can still be generically described by the right-hand side diagram of figure \ref{fig:DIO-collapse}, but with that diagram representing now the three operators of eq.~(\ref{eq:M-NLO}). The coefficients of these operators can be written perturbatively in $\alpha$,
\ali{
\label{eq:Fs:generic}
F_i = F_i^{(\alpha^0)} + \dfrac{\alpha}{4 \pi} F_i^{(\alpha^1)} + \mathcal{O}(\alpha^2),
}
such that
\ali{
\label{eq:Fs:tree}
F_a^{(\alpha^0)} = F_b^{(\alpha^0)} = 0,
\qquad
F_c^{(\alpha^0)} = 1.
}
The virtual corrections, UV-renormalized in on-shell subtraction, read
\bs
\label{eq:Cs-full-theory}
\ali{
F_{a,\rm{OS}}^{(\alpha^1)} &= -1.39732, \\
F_{b,\rm{OS}}^{(\alpha^1)} &= -0.49511, \\
F_{c,\rm{OS}}^{(\alpha^1)} &= \frac{2}{\epsilon} \bigg[ \ln \left(\frac{2 m_{\mu}}{m_e}\right) - 1 \bigg]
- 2 \ln^2 \left(\frac{m_{\mu}}{m_e}\right) + \frac{23}{3} \ln \left(\frac{m_{\mu}}{m_e}\right) - 4 \ln\left(\frac{m_{\mu}}{m_e}\right) \ln\left(\frac{m_e}{\mu}\right) \nonumber \\
& \hspace{30mm} + 4 \ln\left(\frac{m_e}{\mu}\right) + 2 \ln (4) \ln\left(\frac{\mu}{m_{\mu}}\right) +0.10486.
}
\es
We note that $F_{c,\rm{virt.}}^{(\alpha^1)}$ has an IR divergence, regulated by $\epsilon=(4-d)/2$, with $d$ being the dimensions. This is canceled by the one coming from the soft real emission diagrams, depicted in figure \ref{fig:DIO-real}.%
\fn{As in muon conversion, the real emissions from nucleus legs cancel; for details, cf. ref.~\cite{Fontes:2024yvw}.}
These can be calculated via the usual techniques \cite{Denner:2019vbn}.
It is worth stressing that, at one loop, all the integrals of eq.~(\ref{eq:LO-decay-width}) can still be performed in 4 dimensions, and only the integral relative to the real emission must be performed in $d$ dimensions. The final NLO rate is IR finite and given by eq.~(\ref{eq:rate-NLO}).

\bibliographystyle{JHEP}
\bibliography{biblio.bib}

@article{Szafron:2016cbv,
    author = "Szafron, Robert and Czarnecki, Andrzej",
    title = "{Bound muon decay spectrum in the leading logarithmic accuracy}",
    eprint = "1608.05447",
    archivePrefix = "arXiv",
    primaryClass = "hep-ph",
    reportNumber = "ALBERTA-THY-7-16",
    doi = "10.1103/PhysRevD.94.051301",
    journal = "Phys. Rev. D",
    volume = "94",
    number = "5",
    pages = "051301",
    year = "2016"
}

@misc{Pruna:2019CHIPP,
  author       = {Marco Pruna},
  title  = {Low energy Physics/Pheno
            [Contribution to the CHIPP Winter School of Particle Physics, slides 13--15, \url{https://indico.cern.ch/event/744252/contributions/3246688/attachments/1782396/2900341/Pruna_G_M_CHIPP_1_2.pdf}]},
  year         = {2019}
}

@article{Szafron:2015kja,
    author = "Szafron, Robert and Czarnecki, Andrzej",
    title = "{High-energy electrons from the muon decay in orbit: radiative corrections}",
    eprint = "1505.05237",
    archivePrefix = "arXiv",
    primaryClass = "hep-ph",
    reportNumber = "ALBERTA-THY-8-15, FERMILAB-PUB-15-205-PPD",
    doi = "10.1016/j.physletb.2015.12.008",
    journal = "Phys. Lett. B",
    volume = "753",
    pages = "61--64",
    year = "2016"
}

@article{Heeck:2021adh,
    author = "Heeck, Julian and Szafron, Robert and Uesaka, Yuichi",
    title = "{Isotope dependence of muon decay in orbit}",
    eprint = "2110.14667",
    archivePrefix = "arXiv",
    primaryClass = "hep-ph",
    doi = "10.1103/PhysRevD.105.053006",
    journal = "Phys. Rev. D",
    volume = "105",
    number = "5",
    pages = "053006",
    year = "2022"
}

@article{Shtabovenko:2020gxv,
    author = "Shtabovenko, Vladyslav and Mertig, Rolf and Orellana, Frederik",
    title = "{FeynCalc 9.3: New features and improvements}",
    eprint = "2001.04407",
    archivePrefix = "arXiv",
    primaryClass = "hep-ph",
    reportNumber = "P3H-20-002, TTP19-020, TUM-EFT 130/19",
    doi = "10.1016/j.cpc.2020.107478",
    journal = "Comput. Phys. Commun.",
    volume = "256",
    pages = "107478",
    year = "2020"
}

@article{Shtabovenko:2016sxi,
    author = "Shtabovenko, Vladyslav and Mertig, Rolf and Orellana, Frederik",
    title = "{New Developments in FeynCalc 9.0}",
    eprint = "1601.01167",
    archivePrefix = "arXiv",
    primaryClass = "hep-ph",
    reportNumber = "TUM-EFT-71-15",
    doi = "10.1016/j.cpc.2016.06.008",
    journal = "Comput. Phys. Commun.",
    volume = "207",
    pages = "432--444",
    year = "2016"
}

@article{Yennie:1961ad,
    author = "Yennie, D. R. and Frautschi, Steven C. and Suura, H.",
    title = "{The infrared divergence phenomena and high-energy processes}",
    doi = "10.1016/0003-4916(61)90151-8",
    journal = "Annals Phys.",
    volume = "13",
    pages = "379--452",
    year = "1961"
}

@article{Uberall:1960zz,
    author = "Uberall, H.",
    title = "{Decay of mu- Mesons Bound in the K Shell of Light Nuclei}",
    doi = "10.1103/PhysRev.119.365",
    journal = "Phys. Rev.",
    volume = "119",
    pages = "365--376",
    year = "1960"
}

@article{Haenggi:1974hp,
    author = "Haenggi, P. and Viollier, R. D. and Raff, U. and Alder, K.",
    title = "{Muon decay in orbit}",
    doi = "10.1016/0370-2693(74)90193-2",
    journal = "Phys. Lett. B",
    volume = "51",
    pages = "119--122",
    year = "1974"
}

@article{Hoang:2015iva,
    author = "Hoang, Andre H. and Pietrulewicz, Piotr and Samitz, Daniel",
    title = "{Variable Flavor Number Scheme for Final State Jets in DIS}",
    eprint = "1508.04323",
    archivePrefix = "arXiv",
    primaryClass = "hep-ph",
    reportNumber = "DESY-15-098, UWTHPH-2015-19",
    doi = "10.1103/PhysRevD.93.034034",
    journal = "Phys. Rev. D",
    volume = "93",
    number = "3",
    pages = "034034",
    year = "2016"
}

@article{Bundgen:2025utt,
    author = {B{\"u}ndgen, Lars and Harlander, Robert V. and Klein, Sven Yannick and Schaaf, Magnus C.},
    title = "{FeynGame 3.0}",
    eprint = "2501.04651",
    archivePrefix = "arXiv",
    primaryClass = "hep-ph",
    reportNumber = "TTK-24-56, P3H-24-096",
    doi = "10.1016/j.cpc.2025.109662",
    journal = "Comput. Phys. Commun.",
    volume = "314",
    pages = "109662",
    year = "2025"
}

@article{Mertig:1990an,
    author = "Mertig, R. and Bohm, M. and Denner, Ansgar",
    title = "{FEYN CALC: Computer algebraic calculation of Feynman amplitudes}",
    reportNumber = "PRINT-90-0639 (WURZBURG)",
    doi = "10.1016/0010-4655(91)90130-D",
    journal = "Comput. Phys. Commun.",
    volume = "64",
    pages = "345--359",
    year = "1991"
}

@article{Manohar:1997qy,
    author = "Manohar, Aneesh V.",
    title = "{The HQET / NRQCD Lagrangian to order $\alpha/m^3$}",
    eprint = "hep-ph/9701294",
    archivePrefix = "arXiv",
    reportNumber = "UCSD-PTH-97-01",
    doi = "10.1103/PhysRevD.56.230",
    journal = "Phys. Rev. D",
    volume = "56",
    pages = "230--237",
    year = "1997"
}

@article{Isgur:1990yhj,
    author = "Isgur, Nathan and Wise, Mark B.",
    title = "{Weak transition form factors between heavy mesons}",
    reportNumber = "UTPT-90-01, CALT-68-1608",
    doi = "10.1016/0370-2693(90)91219-2",
    journal = "Phys. Lett. B",
    volume = "237",
    pages = "527--530",
    year = "1990"
}

@article{Isgur:1989vq,
    author = "Isgur, Nathan and Wise, Mark B.",
    title = "{Weak Decays of Heavy Mesons in the Static Quark Approximation}",
    reportNumber = "UTPT-89-27, CALT-68-1585",
    doi = "10.1016/0370-2693(89)90566-2",
    journal = "Phys. Lett. B",
    volume = "232",
    pages = "113--117",
    year = "1989"
}

@article{Furry:1951zz,
    author = "Furry, W. H.",
    title = "{On Bound States and Scattering in Positron Theory}",
    doi = "10.1103/PhysRev.81.915",
    journal = "Phys. Rev.",
    volume = "81",
    pages = "115--124",
    year = "1951"
}

@inproceedings{Mu2e-II:2022blh,
    author = "Byrum, K. and others",
    collaboration = "Mu2e-II",
    title = "{Mu2e-II: Muon to electron conversion with PIP-II}",
    booktitle = "{Snowmass 2021}",
    eprint = "2203.07569",
    archivePrefix = "arXiv",
    primaryClass = "hep-ex",
    reportNumber = "FERMILAB-CONF-22-123",
    month = "3",
    year = "2022"
}

@article{COMET:2018auw,
    author = "Abramishvili, R. and others",
    collaboration = "COMET",
    title = "{COMET Phase-I Technical Design Report}",
    eprint = "1812.09018",
    archivePrefix = "arXiv",
    primaryClass = "physics.ins-det",
    doi = "10.1093/ptep/ptz125",
    journal = "PTEP",
    volume = "2020",
    number = "3",
    pages = "033C01",
    year = "2020"
}

@article{Fujii:2023vgo,
    author = "Fujii, Yuki",
    collaboration = "COMET",
    title = "{A search for a muon to electron conversion in COMET}",
    eprint = "2308.14275",
    archivePrefix = "arXiv",
    primaryClass = "hep-ex",
    doi = "10.1088/1748-0221/18/10/C10010",
    journal = "JINST",
    volume = "18",
    number = "10",
    pages = "C10010",
    year = "2023"
}

@article{Mu2e:2014fns,
    author = "Bartoszek, L. and others",
    collaboration = "Mu2e",
    title = "{Mu2e Technical Design Report}",
    eprint = "1501.05241",
    archivePrefix = "arXiv",
    primaryClass = "physics.ins-det",
    reportNumber = "FERMILAB-TM-2594, FERMILAB-DESIGN-2014-01",
    doi = "10.2172/1172555",
    month = "10",
    year = "2014"
}

@article{Diociaiuti:2024stz,
    author = "Diociaiuti, Eleonora",
    collaboration = "Mu2e",
    title = "{Status and perspectives of cLFV at Mu2e}",
    doi = "10.22323/1.457.0017",
    journal = "PoS",
    volume = "WIFAI2023",
    pages = "017",
    year = "2024"
}

@article{SINDRUMII:2006dvw,
    author = "Bertl, Wilhelm H. and others",
    collaboration = "SINDRUM II",
    title = "{A Search for muon to electron conversion in muonic gold}",
    doi = "10.1140/epjc/s2006-02582-x",
    journal = "Eur. Phys. J. C",
    volume = "47",
    pages = "337--346",
    year = "2006"
}

@article{Czarnecki:2011mx,
    author = "Czarnecki, Andrzej and Garcia i Tormo, Xavier and Marciano, William J.",
    title = "{Muon decay in orbit: spectrum of high-energy electrons}",
    eprint = "1106.4756",
    archivePrefix = "arXiv",
    primaryClass = "hep-ph",
    reportNumber = "ALBERTA-THY-08-11",
    doi = "10.1103/PhysRevD.84.013006",
    journal = "Phys. Rev. D",
    volume = "84",
    pages = "013006",
    year = "2011"
}

@article{Fontes:2024yvw,
    author = "Fontes, Duarte and Szafron, Robert",
    title = "{An effective field theory for muon conversion and muon decay-in-orbit}",
    eprint = "2412.05702",
    archivePrefix = "arXiv",
    primaryClass = "hep-ph",
    doi = "10.1007/JHEP05(2025)171",
    journal = "JHEP",
    volume = "05",
    pages = "171",
    year = "2025"
}

@article{Watanabe:1993emp,
    author = "Watanabe, R. and Muto, K. and Oda, T. and Niwa, T. and Ohtsubo, H. and Morita, R. and Morita, M.",
    title = "{Asymmetry and Energy Spectrum of Electrons in Bound-Muon Decay}",
    doi = "10.1006/adnd.1993.1012",
    journal = "Atom. Data Nucl. Data Tabl.",
    volume = "54",
    pages = "165--178",
    year = "1993"
}

@article{Watanabe:1987su,
    author = "Watanabe, R. and Fukui, M. and Ohtsubo, H. and Morita, M.",
    title = "{Angular Distribution of Electrons in Bound Muon Decay}",
    doi = "10.1143/PTP.78.114",
    journal = "Prog. Theor. Phys.",
    volume = "78",
    pages = "114--122",
    year = "1987"
}

@article{Shanker:1979ap,
    author = "Shanker, Oruganti U.",
    title = "{$Z$ Dependence of Coherent $\mu e$ Conversion Rate in Anomalous Neutrinoless Muon Capture}",
    reportNumber = "COO-3066-123",
    doi = "10.1103/PhysRevD.20.1608",
    journal = "Phys. Rev. D",
    volume = "20",
    pages = "1608",
    year = "1979"
}

@article{Shanker:1981mi,
    author = "Shanker, Oruganti U.",
    title = "{High-energy Electrons From Bound Muon Decay}",
    reportNumber = "TRI-PP-81-26",
    doi = "10.1103/PhysRevD.25.1847",
    journal = "Phys. Rev. D",
    volume = "25",
    pages = "1847",
    year = "1982"
}

@article{Shanker:1996rz,
    author = "Shanker, Oruganti U. and Roy, Rajat",
    title = "{High-energy electrons from bound muon decay}",
    reportNumber = "PRINT-97-110 (BOMBAY)",
    doi = "10.1103/PhysRevD.55.7307",
    journal = "Phys. Rev. D",
    volume = "55",
    pages = "7307--7308",
    year = "1997"
}

@article{ParticleDataGroup:2024cfk,
    author = "Navas, S. and others",
    collaboration = "Particle Data Group",
    title = "{Review of particle physics}",
    doi = "10.1103/PhysRevD.110.030001",
    journal = "Phys. Rev. D",
    volume = "110",
    number = "3",
    pages = "030001",
    year = "2024"
}

@article{Kaygorodov:2025yag,
    author = "Kaygorodov, M. Y. and Kozhedub, Y. S. and Malyshev, A. V. and Davydov, A. O. and Wu, Y. and Zhang, S. B.",
    title = "{Study of atomic effects on electron spectrum in bound-muon decay process}",
    eprint = "2506.02416",
    archivePrefix = "arXiv",
    primaryClass = "physics.atom-ph",
    month = "6",
    year = "2025"
}

@article{Beneke:1999br,
    author = "Beneke, M. and Buchalla, G. and Neubert, M. and Sachrajda, Christopher T.",
    title = "{QCD factorization for $B \to \pi \pi$ decays: Strong phases and CP violation in the heavy quark limit}",
    eprint = "hep-ph/9905312",
    archivePrefix = "arXiv",
    reportNumber = "SLAC-PUB-8146, CERN-TH-99-126, SHEP-99-04",
    doi = "10.1103/PhysRevLett.83.1914",
    journal = "Phys. Rev. Lett.",
    volume = "83",
    pages = "1914--1917",
    year = "1999"
}

@article{Beneke:2000ry,
    author = "Beneke, M. and Buchalla, G. and Neubert, M. and Sachrajda, Christopher T.",
    title = "{QCD factorization for exclusive, nonleptonic B meson decays: General arguments and the case of heavy light final states}",
    eprint = "hep-ph/0006124",
    archivePrefix = "arXiv",
    reportNumber = "CERN-TH-2000-159, CLNS-00-1675, PITHA-00-06, SHEP-00-06",
    doi = "10.1016/S0550-3213(00)00559-9",
    journal = "Nucl. Phys. B",
    volume = "591",
    pages = "313--418",
    year = "2000"
}

@article{Szafron:2017guu,
    author = "Szafron, Robert",
    title = "{Radiative Corrections in Bound States}",
    doi = "10.5506/APhysPolB.48.2183",
    journal = "Acta Phys. Polon. B",
    volume = "48",
    pages = "2183",
    year = "2017"
}

@article{Beneke:2015wfa,
    author = "Beneke, M.",
    editor = {Bl\"umlein, Johannes and Jansen, Karl and Kr\"amer, Michael and K\"uhn, Johann H.},
    title = "{Soft-collinear factorization in B decays}",
    eprint = "1501.07374",
    archivePrefix = "arXiv",
    primaryClass = "hep-ph",
    reportNumber = "TUM-HEP-978-15, SFB-CPP-14-119",
    doi = "10.1016/j.nuclphysbps.2015.03.021",
    journal = "Nucl. Part. Phys. Proc.",
    volume = "261-262",
    pages = "311--337",
    year = "2015"
}

@article{Beneke:2005gs,
    author = "Beneke, M. and Yang, D.",
    title = "{Heavy-to-light B meson form-factors at large recoil energy: Spectator-scattering corrections}",
    eprint = "hep-ph/0508250",
    archivePrefix = "arXiv",
    reportNumber = "PITHA-05-10",
    doi = "10.1016/j.nuclphysb.2005.11.027",
    journal = "Nucl. Phys. B",
    volume = "736",
    pages = "34--81",
    year = "2006"
}

@article{Neubert:2004dr,
    author = "Neubert, M.",
    editor = "Blaschke, D. and Ivanov, M. A. and Mannel, T.",
    title = "{Theory of exclusive hadronic B decays}",
    doi = "10.1007/978-3-540-40975-5_1",
    journal = "Lect. Notes Phys.",
    volume = "647",
    pages = "3--41",
    year = "2004"
}

@article{Beneke:2003pa,
    author = "Beneke, M. and Feldmann, T.",
    title = "{Factorization of heavy to light form-factors in soft collinear effective theory}",
    eprint = "hep-ph/0311335",
    archivePrefix = "arXiv",
    reportNumber = "PITHA-03-11, CERN-TH-2003-286",
    doi = "10.1016/j.nuclphysb.2004.02.033",
    journal = "Nucl. Phys. B",
    volume = "685",
    pages = "249--296",
    year = "2004"
}

@book{Berestetskii:1982qgu,
    author = "Berestetskii, V. B. and Lifshitz, E. M. and Pitaevskii, L. P.",
    title = "{Quantum Electrodynamics}",
    isbn = "978-0-7506-3371-0",
    publisher = "Pergamon Press",
    address = "Oxford",
    series = "Course of Theoretical Physics",
    volume = "4",
    year = "1982"
}

@article{Harlander:2020cyh,
    author = "Harlander, R. V. and Klein, S. Y. and Lipp, M.",
    title = "{FeynGame}",
    eprint = "2003.00896",
    archivePrefix = "arXiv",
    primaryClass = "physics.ed-ph",
    reportNumber = "TTK-20-04",
    doi = "10.1016/j.cpc.2020.107465",
    journal = "Comput. Phys. Commun.",
    volume = "256",
    pages = "107465",
    year = "2020"
}

@article{Harlander:2024qbn,
    author = "Harlander, Robert and Klein, Sven Yannick and Schaaf, Magnus C.",
    title = "{FeynGame-2.1 -- Feynman diagrams made easy}",
    eprint = "2401.12778",
    archivePrefix = "arXiv",
    primaryClass = "hep-ph",
    reportNumber = "TTK-24-05",
    doi = "10.22323/1.449.0657",
    journal = "PoS",
    volume = "EPS-HEP2023",
    pages = "657",
    year = "2024"
}

@article{Denner:2019vbn,
    author = "Denner, Ansgar and Dittmaier, Stefan",
    title = "{Electroweak Radiative Corrections for Collider Physics}",
    eprint = "1912.06823",
    archivePrefix = "arXiv",
    primaryClass = "hep-ph",
    reportNumber = "FR-PHENO-019",
    doi = "10.1016/j.physrep.2020.04.001",
    journal = "Phys. Rept.",
    volume = "864",
    pages = "1--163",
    year = "2020"
}

@article{Hill:2023bfh,
    author = "Hill, Richard J. and Plestid, Ryan",
    title = "{All orders factorization and the Coulomb problem}",
    eprint = "2309.15929",
    archivePrefix = "arXiv",
    primaryClass = "hep-ph",
    reportNumber = "CALT-TH-2023-034, FERMILAB-PUB-23-454-T",
    doi = "10.1103/PhysRevD.109.056006",
    journal = "Phys. Rev. D",
    volume = "109",
    number = "5",
    pages = "056006",
    year = "2024"
}

@article{Hill:2023acw,
    author = "Hill, Richard J. and Plestid, Ryan",
    title = "{Field Theory of the Fermi Function}",
    eprint = "2309.07343",
    archivePrefix = "arXiv",
    primaryClass = "hep-ph",
    reportNumber = "CALT-TH/2023-029, FERMILAB-PUB-23-453-T",
    doi = "10.1103/PhysRevLett.133.021803",
    journal = "Phys. Rev. Lett.",
    volume = "133",
    number = "2",
    pages = "021803",
    year = "2024"
}

@article{Neubert:1993mb,
    author = "Neubert, Matthias",
    title = "{Heavy quark symmetry}",
    eprint = "hep-ph/9306320",
    archivePrefix = "arXiv",
    reportNumber = "SLAC-PUB-6263",
    doi = "10.1016/0370-1573(94)90091-4",
    journal = "Phys. Rept.",
    volume = "245",
    pages = "259--396",
    year = "1994"
}

@book{Manohar:2000dt,
    author = "Manohar, Aneesh V. and Wise, Mark B.",
    title = "{Heavy quark physics}",
    isbn = "978-0-521-03757-0",
    volume = "10",
    year = "2000",
    publisher = "Cambridge University Press"
}

@article{Caswell:1985ui,
    author = "Caswell, W. E. and Lepage, G. P.",
    title = "{Effective Lagrangians for Bound State Problems in QED, QCD, and Other Field Theories}",
    reportNumber = "CLNS-85/641",
    doi = "10.1016/0370-2693(86)91297-9",
    journal = "Phys. Lett. B",
    volume = "167",
    pages = "437--442",
    year = "1986"
}

@article{Kinoshita:1995mt,
    author = "Kinoshita, T. and Nio, M.",
    title = "{Radiative corrections to the muonium hyperfine structure. I. The $\alpha^2 (Z \alpha)$ correction}",
    eprint = "hep-ph/9512327",
    archivePrefix = "arXiv",
    reportNumber = "CLNS-95-1382",
    doi = "10.1103/PhysRevD.53.4909",
    journal = "Phys. Rev. D",
    volume = "53",
    pages = "4909--4929",
    year = "1996"
}

@article{Paz:2015uga,
    author = "Paz, Gil",
    title = "{An Introduction to NRQED}",
    eprint = "1503.07216",
    archivePrefix = "arXiv",
    primaryClass = "hep-ph",
    reportNumber = "WSU-HEP-1502",
    doi = "10.1142/S021773231550128X",
    journal = "Mod. Phys. Lett. A",
    volume = "30",
    number = "26",
    pages = "1550128",
    year = "2015"
}

@article{Pineda:1997bj,
    author = "Pineda, A. and Soto, J.",
    editor = "Narison, Stephan",
    title = "{Effective field theory for ultrasoft momenta in NRQCD and NRQED}",
    eprint = "hep-ph/9707481",
    archivePrefix = "arXiv",
    reportNumber = "UB-ECM-PF-97-17",
    doi = "10.1016/S0920-5632(97)01102-X",
    journal = "Nucl. Phys. B Proc. Suppl.",
    volume = "64",
    pages = "428--432",
    year = "1998"
}

@article{Pineda:1997ie,
    author = "Pineda, Antonio and Soto, Joan",
    title = "{The Lamb shift in dimensional regularization}",
    eprint = "hep-ph/9711292",
    archivePrefix = "arXiv",
    reportNumber = "UB-ECM-PF-97-33",
    doi = "10.1016/S0370-2693(97)01537-2",
    journal = "Phys. Lett. B",
    volume = "420",
    pages = "391--396",
    year = "1998"
}

@article{Brambilla:1999xf,
    author = "Brambilla, Nora and Pineda, Antonio and Soto, Joan and Vairo, Antonio",
    title = "{Potential NRQCD: An Effective theory for heavy quarkonium}",
    eprint = "hep-ph/9907240",
    archivePrefix = "arXiv",
    reportNumber = "CERN-TH-99-199, HEPHY-PUB-716-99, UB-ECM-PF-99-06, UWTHPH-1999-34, UB-ECM-PF-99-13",
    doi = "10.1016/S0550-3213(99)00693-8",
    journal = "Nucl. Phys. B",
    volume = "566",
    pages = "275",
    year = "2000"
}

@article{Beneke:1999qg,
    author = "Beneke, M. and Signer, A. and Smirnov, Vladimir A.",
    title = "{Top quark production near threshold and the top quark mass}",
    eprint = "hep-ph/9903260",
    archivePrefix = "arXiv",
    reportNumber = "CERN-TH-99-57, DTP-99-26",
    doi = "10.1016/S0370-2693(99)00343-3",
    journal = "Phys. Lett. B",
    volume = "454",
    pages = "137--146",
    year = "1999"
}

@inproceedings{Beneke:1998jj,
    author = "Beneke, M.",
    title = "{New results on heavy quarks near threshold}",
    booktitle = "{3rd Workshop on Continuous Advances in QCD (QCD 98)}",
    eprint = "hep-ph/9806429",
    archivePrefix = "arXiv",
    reportNumber = "CERN-TH-98-202",
    pages = "293--309",
    month = "6",
    year = "1998"
}

@article{Bauer:2000ew,
    author = "Bauer, Christian W. and Fleming, Sean and Luke, Michael E.",
    title = "{Summing Sudakov logarithms in $B \to  X_s \gamma$ in effective field theory.}",
    eprint = "hep-ph/0005275",
    archivePrefix = "arXiv",
    reportNumber = "UTPT-00-03",
    doi = "10.1103/PhysRevD.63.014006",
    journal = "Phys. Rev. D",
    volume = "63",
    pages = "014006",
    year = "2000"
}

@article{Bauer:2000yr,
    author = "Bauer, Christian W. and Fleming, Sean and Pirjol, Dan and Stewart, Iain W.",
    title = "{An Effective field theory for collinear and soft gluons: Heavy to light decays}",
    eprint = "hep-ph/0011336",
    archivePrefix = "arXiv",
    reportNumber = "UCSD-PTH-00-28",
    doi = "10.1103/PhysRevD.63.114020",
    journal = "Phys. Rev. D",
    volume = "63",
    pages = "114020",
    year = "2001"
}

@article{Bauer:2001ct,
    author = "Bauer, Christian W. and Stewart, Iain W.",
    title = "{Invariant operators in collinear effective theory}",
    eprint = "hep-ph/0107001",
    archivePrefix = "arXiv",
    reportNumber = "UCSD-PTH-01-09",
    doi = "10.1016/S0370-2693(01)00902-9",
    journal = "Phys. Lett. B",
    volume = "516",
    pages = "134--142",
    year = "2001"
}

@article{Bauer:2001yt,
    author = "Bauer, Christian W. and Pirjol, Dan and Stewart, Iain W.",
    title = "{Soft collinear factorization in effective field theory}",
    eprint = "hep-ph/0109045",
    archivePrefix = "arXiv",
    reportNumber = "UCSD-PTH-01-15",
    doi = "10.1103/PhysRevD.65.054022",
    journal = "Phys. Rev. D",
    volume = "65",
    pages = "054022",
    year = "2002"
}

@article{Beneke:2002ph,
    author = "Beneke, M. and Chapovsky, A. P. and Diehl, M. and Feldmann, T.",
    title = "{Soft collinear effective theory and heavy to light currents beyond leading power}",
    eprint = "hep-ph/0206152",
    archivePrefix = "arXiv",
    reportNumber = "PITHA-02-09",
    doi = "10.1016/S0550-3213(02)00687-9",
    journal = "Nucl. Phys. B",
    volume = "643",
    pages = "431--476",
    year = "2002"
}

@article{Beneke:2002ni,
    author = "Beneke, M. and Feldmann, T.",
    title = "{Multipole expanded soft collinear effective theory with nonAbelian gauge symmetry}",
    eprint = "hep-ph/0211358",
    archivePrefix = "arXiv",
    reportNumber = "PITHA-02-17",
    doi = "10.1016/S0370-2693(02)03204-5",
    journal = "Phys. Lett. B",
    volume = "553",
    pages = "267--276",
    year = "2003"
}

@article{Fleming:2007qr,
    author = "Fleming, Sean and Hoang, Andre H. and Mantry, Sonny and Stewart, Iain W.",
    title = "{Jets from massive unstable particles: Top-mass determination}",
    eprint = "hep-ph/0703207",
    archivePrefix = "arXiv",
    reportNumber = "MIT-CTP-3791, CALT-68-2624, MPP-2007-9",
    doi = "10.1103/PhysRevD.77.074010",
    journal = "Phys. Rev. D",
    volume = "77",
    pages = "074010",
    year = "2008"
}

@article{Fleming:2007xt,
    author = "Fleming, Sean and Hoang, Andre H. and Mantry, Sonny and Stewart, Iain W.",
    title = "{Top Jets in the Peak Region: Factorization Analysis with NLL Resummation}",
    eprint = "0711.2079",
    archivePrefix = "arXiv",
    primaryClass = "hep-ph",
    reportNumber = "MIT-CTP-3868, CALT-68-2625, MPP-2007-103",
    doi = "10.1103/PhysRevD.77.114003",
    journal = "Phys. Rev. D",
    volume = "77",
    pages = "114003",
    year = "2008"
}

\end{document}